\documentclass[letterpaper,twocolumn,10pt]{article}
\usepackage{usenix-2020-09}

\usepackage{tikz}
\usepackage{amsmath}

\usepackage{filecontents}
\usepackage{hyperref} 

\usepackage{amssymb}
\usepackage{graphicx}
\usepackage{subfig}
\usepackage{mathtools}
\usepackage{amsthm}
\usepackage{algorithm}
\usepackage{algorithmic}
\usepackage{xcolor}
\usepackage{newtxmath}
\usepackage{url}
\usepackage{tablefootnote}
\usepackage{multirow}
\usepackage{url}
\usepackage{makecell}
\usepackage{changepage}
\usepackage{amsthm}
\usepackage{tikz}
\usepackage{mwe}
\usepackage[export]{adjustbox}
\newtheorem{assumption}{Assumption}
\usepackage{graphics}
\usepackage{makecell}
\usepackage{multirow}
\usepackage{comment}
\usepackage{tablefootnote}
\usepackage{pifont}
\usepackage{scalerel}
\theoremstyle{definition}
\newtheorem{theorem}{Theorem}[section]
\newtheorem{definition}{Definition}[section]

\newcommand{\Bigram}{\mathsf{Bigram}}

\newcommand{\revise}[1]{{\color{blue}#1}}
\renewcommand{\revise}[1]{{#1}}
\usepackage{titlesec}
\begin{document}
%----------------------------------------------------------------------------

\title{Provably Robust Multi-bit Watermarking for AI-generated Text } % TODO: replace with your title

 \author{
\rm{ Wenjie Qu$^{1}$, 
Wengrui Zheng$^{1}$, 
Tianyang Tao$^{1}$, 
Dong Yin$^{1}$, 
Yanze Jiang$^{1}$, 
} \\
\rm{Zhihua Tian$^{1}$, Wei Zou$^{2}$, 
Jinyuan Jia$^{2}$, 
Jiaheng Zhang$^{1}$} \\
{ $^{1}$National University of Singapore, $^{2}$Pennsylvania State University}
}

% make the title area
\maketitle

\begin{abstract}
Large Language Models (LLMs) have demonstrated remarkable capabilities of generating texts resembling human language. However, they can be misused by criminals to create deceptive content, such as fake news and phishing emails, which raises ethical concerns. Watermarking is a key technique to address these concerns, which embeds a message (e.g., a bit string) into a text generated by an LLM. 
By embedding the user ID (represented as a bit string) into generated texts, we can trace generated texts to the user, known as content source tracing.
The major limitation of existing watermarking techniques is that they achieve sub-optimal performance for content source tracing in real-world scenarios. The reason is that they cannot accurately or efficiently extract a long message from a generated text. We aim to address the limitations. 

In this work, we introduce a new watermarking method for LLM-generated text grounded in pseudo-random segment assignment.  We also propose multiple techniques to further enhance the robustness of our watermarking algorithm.
We conduct extensive experiments to evaluate our method.
Our experimental results show that our method \revise{achieves  a much better tradeoff between extraction accuracy and time complexity, compared with existing baselines}. 
For instance, when embedding a message of length 20 into a 200-token generated text, our method achieves a match rate of $97.6\%$, while the state-of-the-art work Yoo et al. only achieves $49.2\%$. Additionally, we prove that our watermark can tolerate edits within an edit distance of 17 on average for each paragraph under the same setting.

 \end{abstract}

\section{Introduction}
\label{intro}

\begin{figure}[!t]
\centering
\subfloat{\includegraphics[width=0.95\linewidth]{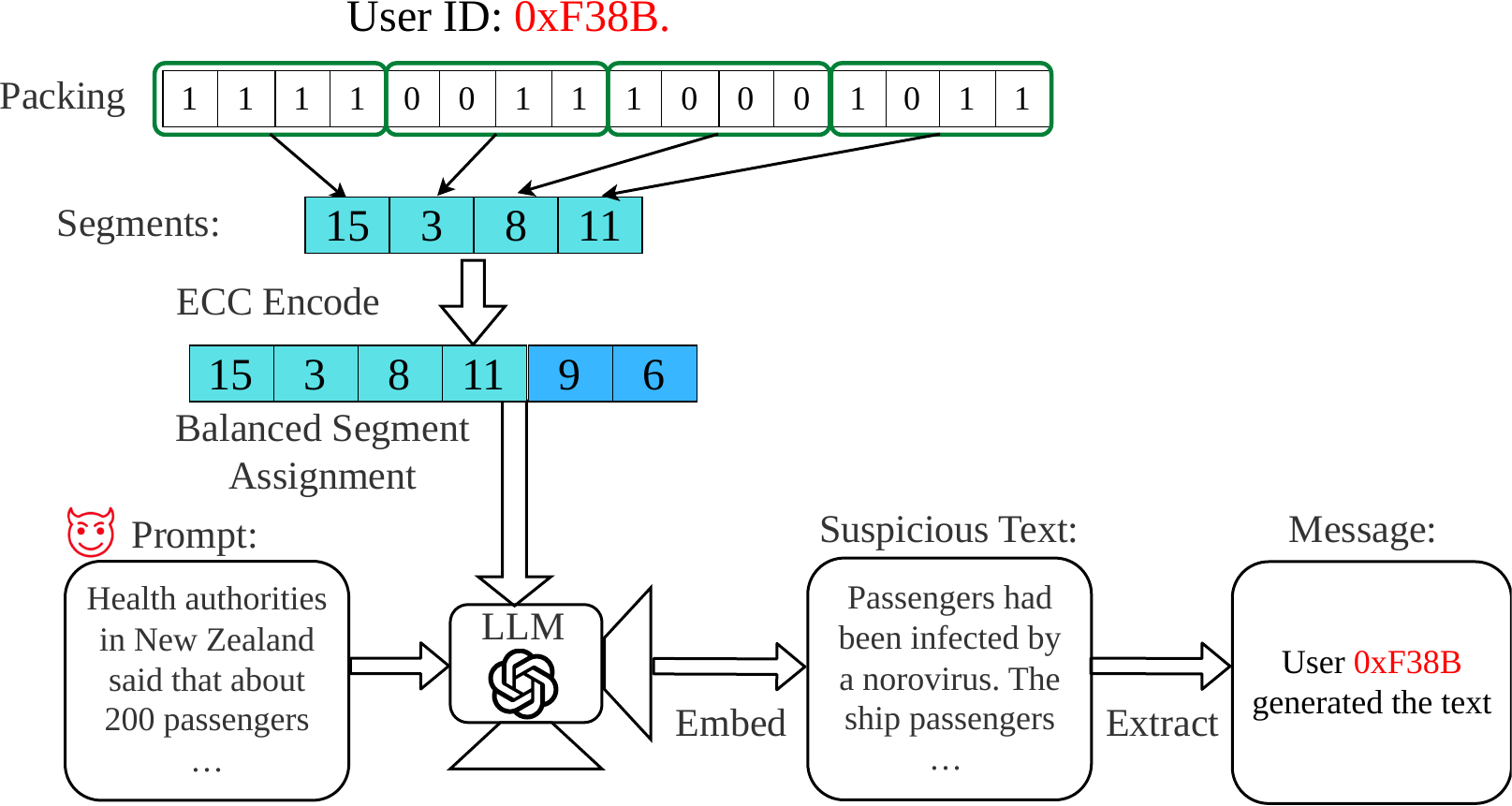}}
\caption{Outline of our watermarking application scenario and workflow. Users query the LLM with a prompt. During text generation, using our watermarking method, the service provider embeds a unique user ID into the generated text. Later, when some suspicious LLM-generated text used for malicious purposes is found, the service provider can identify and extract the watermark to trace the original user who generated the text.}
\label{fig:concept}
\vspace{-3mm}
\end{figure}

Generative models such as GPT-4~\cite{openai2023gpt4}, Stable Diffusion~\cite{rombach2021highresolution}, Wavenet~\cite{oord2016wavenet}, VideoPoet~\cite{kondratyuk2023videopoet}  could generate remarkably human-like content such as text, image, audio, video, etc.. Consequently, they are utilized in various real-world applications such as ChatBot~\cite{openai2023chatgpt}, Web searching~\cite{microsoft2023newbing}, and programming assistance~\cite{github2023copilot}. Despite their fascinating capabilities, LLMs could also be misused to generate fake news~\cite{9534192}, phishing emails~\cite{karanjai2022targeted}, fake product reviews~\cite{adelani2020generating}, resulting in severe ethical concerns for their deployment in the real world. Watermarking~\cite{cox2002digital} is a key technique to address those concerns. It enables the detection and tracing of machine-generated content to specific users, ensuring authenticity and deterring misuse. Generally speaking, a watermarking method consists of two functions: \emph{embedding} and \emph{extraction}. The embedding function embeds a piece of pre-selected message (usually a bit string) into generated content (e.g., image, text, audio, video), while the extraction function identifies the watermark and extracts the message embedded in the content.

%Different watermarking methods essentially design 

%the model provider also knows which user generates the given text using its model .

% With the recent emergence of Large Language Models(LLMs), such as GPT-4, their capability of generating highly realistic human-like text raises concerns about misusing these models to create deceptive content. LLMs can generate fake news\cite{9534192}, 

%Watermarking has various usages in auditing generative AI, including improving accountability, transparency,  and protecting copyright.   It has been well-studied in domains like image~\cite{yeung1997invisible,tang2003feature,zhang2020model} and audio~\cite{swanson1998robust,bassia2001robust,zhang2022robust}.  However, it remains less explored in the text domain.

%Watermarking has various usages in auditing LLMs, including improving accountability, transparency,  and protecting copyright. 

In this paper, we consider typical usage scenarios for watermarking LLM-generated texts. Suppose a model provider deploys an LLM (e.g., GPT-4~\cite{openai2023gpt4}) as a cloud service.  When a user queries the LLM with a prompt, the watermarking algorithm enables the service provider to embed a unique user ID into the generated text. Given a text that is suspected to be LLM-generated (e.g., fake news on social media), the watermark extraction function can be used to identify whether the text is watermarked. Furthermore, if the watermark is identified, the user ID can be extracted from the watermark to trace the original user who uses the LLM to generate the given text.

\noindent
\textbf{Limitations of existing text watermarking methods.}
 Most of the existing studies in text watermarking~\cite{pmlr-v202-kirchenbauer23a,christ2023undetectable,kuditipudi2024robust,lee2023wrote,zhao2023provable,piet2023mark,fairoze2023publicly,dathathri2024scalable,wuresilient,giboulot2024watermax,hou2023semstamp,munyer2023deeptextmark} can only be used to detect whether a text is generated by an LLM or not. However, they are not suitable for use under the previously mentioned scenario of content source tracing. This stems from the fact that they only embed a single indicator that the text is watermarked. In response, several pioneering works~\cite{fernandez2023three,wang2023towards} have explored the design of multi-bit watermarking schemes for LLMs. Nevertheless, their deployments in real-world scenarios are hindered by the high computational cost of their extraction function. For example,  \cite{fernandez2023three} requires about 8 hours to extract a 32-bit message from a given text. In other words, these approaches lack scalability when dealing with long messages, posing significant challenges for practical deployments. \revise{~\cite{yoo2023advancing} proposed to pseudo-randomly allocate short bit segments (1-3 bits) to enhance extraction efficiency.} However, their method falls short in the accuracy of extracting the embedded message from watermarked texts. For example, when the message length is 16 bits, their method can only extract the embedded message from the watermarked text with a probability of 73.6\%. In essence, a scalable text watermarking method that can accurately and efficiently embed and extract multi-bit messages from generated text remains elusive.

\noindent
\textbf{Goals in designing practical watermarking methods.} Due to the requirements of content source tracing, there are four key goals when designing a watermarking method for texts: \emph{multi-bit capacity}, \emph{correctness}, \emph{robustness} and \emph{efficiency}. 

\begin{itemize}
    \item \textbf{Multi-bit Capacity:}
A watermarking method should allow embedding a multi-bit message into a generated text. This goal is essential to content source tracing~\cite{lubin2003robust}. 
%For instance, when a user uses LLM services (e.g. OpenAI), the model provider  should be able to embed the user's ID into his generated text. Given a piece of watermarked text, the model provider should be able to know which user generated that text.\jiaheng{If no space, we can delete the example in the def. }

\item \textbf{Correctness:} A watermarking method should enable the embedded message to be correctly extracted from a watermarked text.
%i.e., the embedded message should be correctly extracted. 

\item \textbf{Robustness:} A watermarking method should be robust against post-processing operations such as word deletion, synonym substitution, and paraphrasing. This goal ensures the embedded message can be reliably extracted even if the watermarked text undergoes various forms of editing or alteration.
%, maintaining the integrity of the watermark. %\jiaheng{Bounded or limited.}

\item  \textbf{Efficiency:} The efficiency goal requires the computational overhead of the watermarking method during both the embedding and extraction process to be small, which is essential for real-world usage.
\end{itemize}

\noindent
\textbf{Our work.} This work focuses on designing a watermarking scheme that simultaneously meets the four goals. Designing such a scheme is particularly challenging because the information redundancy in discrete text is much more limited compared to that in images or videos. To ensure a high probability of correct extraction, our critical insight is to allocate the information redundancy evenly across each bit of the message.

%In this work, we address this challenge and successfully achieve the four necessary properties of text watermarking simultaneously for the first time. 

\revise{To achieve an even distribution of information redundancy without losing efficiency, we design an approach unifying both message enumeration-based methods and bit assignment-based methods.  We partition the entire message into multiple \textbf{bit segments} for embedding. During each token's generation, we pseudo-randomly assign a segment to the token and embed it leveraging ideas in enumeration-based watermarks. 
Our design makes the information redundancy distributed to each bit more balanced while avoiding the overhead of enumerating all possible messages during extraction.  Our method offers a systematic way to balance the trade-off between watermark accuracy and computational efficiency.}

We also design two techniques to refine our method:
\begin{itemize}
    \item \noindent \textbf{Balanced segment assignment.} We propose a strategy to further eliminate the imbalance in pseudo-random segment assignment. In particular, we employ dynamic programming to allocate tokens to each segment in a balanced manner. As a result, the probability of extracting erroneous segments is significantly reduced.

    %This strategy employs dynamic programming to efficiently determine the most balanced assignment plan. By allocating tokens to each segment in a balanced manner, the probability of extracting erroneous segments is significantly reduced.

    \item \noindent \textbf{Adopting error-correction code.}
Watermarked text may be edited by users, potentially distorting the embedded message. As a result, the extraction process might yield an incorrect message. To enhance the accuracy and robustness of our method, we utilize error-correction codes (ECC) to encode our segments before embedding them into the text. By leveraging ECC's ability to correct errors in the embedded segments, our watermark can tolerate more extensive edits.
\end{itemize}

%\noindent \textbf{Pairing-based segment embedding.} To improve extraction accuracy, we propose an enhanced method to embed each segment into tokens based on the idea of pairing potential values. During generation of each token, instead of embedding watermark through biasing a random subset of vocabulary, we further optimize this method by increasing the distinguishability among biased vocabulary subsets of different segment values. Through making the biased vocabulary subsets of paired segment values disjoint from each other, our solution further increases the extraction accuracy.

%We also observe that directly applying the techniques from~\cite{fernandez2023three, wang2023towards} to embed each segment into its allocated tokens is suboptimal \jiaheng{since...}. Based on this observation, we propose a method to embed segments . \jiaheng{Explain pairing based method. } This method improves the probability of correct extraction.

Additionally, we derive the provable robustness guarantee of our watermarking method. We prove that our method can reliably extract the correct watermark from a watermarked text as long as the total number of word or token edits (insertion, deletion, substitution) remains within a certain bound.

To validate the performance of our method, we conduct extensive experiments on multiple benchmark datasets (OpenGen~\cite{krishna2023paraphrasing},  C4 news dataset \cite{raffel2020exploring}, Essays dataset \cite{essaysdataset}) with multiple large language models (LLaMA-2-7B~\cite{touvron2023llama}, Guanaco-7B~\cite{dettmers2024qlora}, and Falcon-7B~\cite{almazrouei2023falcon}). We use \textbf{match rate} as the evaluation metric, which measures the proportion of generated texts that can exactly extract embedded message without error. We have the following observations from the experimental results. First, our watermarking method could extract the embedded message from a watermarked text with an extremely high match rate. For instance, our method achieves a $97.6\%$ match rate when the lengths of the message and watermarked text are 20 bits and 200 tokens respectively, while Yoo et al.~\cite{yoo2023advancing} only has a match rate of $49.2\%$.  
Second, our watermark is robust against manipulations. For instance, under the same setting, when the attacker applies copy-paste attack, our method still retains $90\%$ match rate, surpassing the match rate of Yoo et al.~\cite{yoo2023advancing} which drops to $32\%$. 
In addition, we prove that our watermark can tolerate edits within an edit distance of 17 on average for each paragraph under
the same setting.
Third, our method also enjoys remarkable efficiency, capable of extracting a 32-bit message from a sentence in just 0.6 seconds, whereas~\cite{wang2023towards} requires 8,300 seconds. Our results show that our method also maintains the quality of the text. In the experiments, we observe a negligible alteration in the distribution of Perplexity (PPL) between watermarked and unwatermarked text.  These experimental results validate that our scheme simultaneously achieves the four goals.

%\jiaheng{Do we need to mention the high efficiency as we state the weakness of previous works? }

Our key contributions are summarized as follows. 
\begin{itemize}
    \item \revise{We propose a new Large Language Model (LLM) multi-bit watermarking scheme. Our innovation lies in the watermark design which unifies message enumeration-based methods and bit assignment-based methods. We also proposed further improvement techniques. 
    Our scheme offers a systematic approach to balancing the trade-off between watermark accuracy and computation efficiency in multi-bit watermarking algorithm, while simultaneously enhancing the watermark robustness and preserving text quality.}
    %It provides a ground-breaking solution for AI-generated text auditing.

    \item 
    We are the first to derive the theoretical robustness guarantee under text edit distance for LLM multi-bit watermarking based on probabilistic analysis. The robustness bound for each generated paragraph can be efficiently computed in practice. %\jiaheng{Additionally, we explore efficient methods for computing the robustness bound in advance. } 

    %We derive the robustness bound under text edit distance for our proposed LLM watermarking method based on probabilistic analysis.

    \item We validate the effectiveness of our proposed scheme through extensive experiments. Our results demonstrate a significant outperformance compared to existing multi-bit watermarking schemes,  particularly in terms of the correctness of extraction, robustness against different types of attacks, and efficiency of extraction. 
    We release our source code at \url{https://github.com/randomizedtree/segment-watermark}.
\end{itemize}

    %Furthermore, we have carried out an experiment to simulate a real-world application scenario, namely content source tracing. 

\section{Background and Related Work}
\subsection{Zero-bit watermarking}
\label{background-zero-bit-watermark}
 Watermarking digital text is challenging due to the discrete nature of text~\cite{johnson2001information}.   While watermarking on image~\cite{nikolaidis1999digital,potdar2005survey} and other domains focuses on embedding long bit strings, most of the research on text watermarking focuses on zero-bit watermarking, namely only identifying whether the content is watermarked or not. The main application of zero-bit watermarking is distinguishing between machine-generated text (watermarked) and human-written text (unwatermarked).
 
 \noindent \textbf{Existing works on zero-bit watermarking.} Various research has been conducted on zero-bit watermarking in text. Early approaches are mainly rule-based, such as paraphrasing \cite{atallah2002natural} and synonym substitution~\cite{topkara2006hiding}.  Later, advancements in modern language models led to improved methods. In \cite{abdelnabi2021adversarial}, they designed a watermarking framework in which both embedding and extraction are handled by text-to-text adversarially trained language models. \cite{he2022cater,he2022protecting} embeds watermarks by context-aware lexical substitution.  
 Recently, Kirchenbauer et al.~\cite{pmlr-v202-kirchenbauer23a}  proposed imperceptible watermarks by modifying the output logits of language models during token generation. This approach has emerged as a promising approach for distinguishing language model-generated text from human-written text.  ~\cite{christ2023undetectable,kuditipudi2024robust,wuresilient} proposed distortion-free watermarking schemes which retain the original LLM's output distribution. Zhang et al.~\cite{zhang2023remark} trained a message-encoding neural network for watermark injection, followed by a message-decoding neural network for watermark extraction. Zhao et al.~\cite{zhao2023provable} proposed a watermarking construction that offers provable robustness guarantees for editing properties.    \cite{ren2023robust,liu2023semantic} proposed semantic-based watermarks that enhance robustness against paraphrasing for zero-bit watermarking. \revise{ \cite{giboulot2024watermax} formulated watermarking as a search problem that aims to identify the token chunks with the lowest p-value among the top outputs of LLM.}
 
  \noindent \textbf{State-of-the-art zero-bit watermarking~\cite{pmlr-v202-kirchenbauer23a}.} Next, we introduce a state-of-the-art zero-bit watermarking solution for LLM. 
%\noindent
%\textbf{State-of-the-art zero-bit watermarking~\cite{pmlr-v202-kirchenbauer23a}:}
%\label{sec:zero-bit}
The main idea of Kirchenbauer et al.~\cite{pmlr-v202-kirchenbauer23a} 
is to bias a subset of tokens to be more frequently generated during token generation. Suppose we have an LLM with vocabulary $V$. Given a prompt as input, the LLM autoregressively generates a response. At each decoding step $i$ (i.e., the LLM generates the $i$-th token), the LLM's decoded token is sampled from logits vector $v_i$ 
 $ \in \mathbb {R} ^{|V|}$, where $|V|$ represents the size of $V$. For instance, in greedy decoding, the $i$-th output token is $\arg \max_{j} v_{ij}$. 

To embed the watermark, at step $i$, a subset of tokens $G$ is selected from the vocabulary $V$, i.e., $G \subseteq V$. $G$ is called the \textbf{green list}, while $V\setminus G$ is called the \textbf{red list}. The ratio of  $|G|$ to $|V|$ is a fixed hyperparameter.  The green/red list selection is determined by a random seed $s$ obtained by a hash function taking the $(i-1)$-th token as input. The logits vector is modified by adding bias term $\delta$ to all tokens in $G$. The text generated by LLM is watermarked when sampled from the modified logits vector, because the $i$-th output token is more likely to belong to the \textbf{green list} $G$ seeded with the $(i-1)$-th token, than to the red list.

%For example, we can partition the green and red lists to be equally sized, such that $|G|=|V\setminus G|=\frac{1}{2}|V|$.

If the $i$-th output token belongs to the green list $G$ seeded with the $(i$-$1)$-th token, we call it a \textbf{green token}. After generating the text with a watermark, the addition of bias results in the majority of tokens being green tokens. For unwatermarked text, by expectation, if the green list and the red list are equally partitioned, approximately half of the tokens are green tokens and the remaining half of them are red tokens. Therefore, the service provider could leverage this distribution difference through statistical testing, namely Z-test~\cite{casella2021statistical}, to detect whether the text is watermarked or not.

%\subsection{Existing Multi-bit Watermarking and Their Limitations}
\subsection{Multi-bit watermarking}
\label{sec:related_work_multi}
%In this subsection, we discuss existing multi-bit watermarking~\cite{fernandez2023three,wang2023towards,yoo2023advancing,boroujeny2024multi} and their limitations.

%\noindent
%\textbf{Language model multi-bit watermarks.} 
Another line of work in LLM watermarking is multi-bit watermarking. Different from zero-bit watermarking, the embedding function of multi-bit watermarking embeds a multi-bit message into the generated text, whereas its extraction function extracts the multi-bit message from the given text. While the applications of zero-bit watermarking are mainly restricted to detection of LLM-generated text, multi-bit watermarks can be applied in broader scenarios such as content source tracing.

Multiple works~\cite{wang2023towards,fernandez2023three,yoo2023advancing} generalized the idea of selecting green tokens and adding bias to them in~\cite{pmlr-v202-kirchenbauer23a}.
Wang et al.~\cite{wang2023towards} proposed a multi-bit watermarking technique by using the message content as the seed to randomly select green list words.  Fernandez et al.~\cite{fernandez2023three}  consolidated multi-bit watermarks for LLMs
by adopting more robust statistical tests. 
% and associating different secret keys for different messages. 
Concurrent with our work, Boroujeny et al.~\cite{boroujeny2024multi} extended the distortion-free watermark in~\cite{kuditipudi2024robust} to the multi-bit setting \revise{, while \cite{zamir2024excuse} generalized \cite{christ2023undetectable} to achieve multi-bit distortion-free watermarking}. Another concurrent work Cohen et al~\cite{cohenwatermarking} proposed a block-based multi-bit watermarking method.
More similar to our work, Yoo et al.~\cite{yoo2023advancing} proposed to pseudo-randomly assign each token \revise{a short bit segment (1-3 bits)} to embed. 
%This makes the decoding of long information much more efficient, also they can achieve good \textbf{bit accuracy} (ratio of embedded bits extracted that matches the ground truth). However, under most real-world scenarios, service providers are more interested in the exact match of extracted information, not a partial match. And their performance on a more realistic metric, \textbf{match rate}\footnote{the definition is equivalent to packet error ratio in \cite{yoo2023advancing} Appendix A.7} (ratio of text generation that extracted information exactly match groundtruth), is still unsatisfying as shown in our results. 
Next, we provide a detailed discussion on the limitations of the existing multi-bit watermarking schemes. We also discuss the differences between our work and other works in LLM watermarking which also leverage error-correction code.
%,  due to lack of error-correction.

\noindent
\textbf{Limitations of existing message enumeration-based multi-bit watermark.} 
%Recall Section~\ref{sec:zero-bit}, in zero-bit watermarking~\cite{pmlr-v202-kirchenbauer23a}, the hash function takes the previous token as input to compute the random seed for green/red list partition. Hash function key based multi-bit watermark\wenjie{inconsistent name}~\cite{fernandez2023three,wang2023towards} generalizes \cite{pmlr-v202-kirchenbauer23a} by combining the message to embed into the input of the hash function.  Namely, at the $i$-th step of token generation, the hash function takes both information $K$ and the ($i$-$1$)-th token as input.
During embedding, message enumeration based watermarks~\cite{fernandez2023three,wang2023towards} determines the green list by letting the hash function to take both the message $K$ and the previous token as input.
During extraction, the service provider needs to enumerate all possible messages and compute the number of green tokens for each candidate. Then the message candidate that corresponds to the maximal green tokens is regarded as the message embedded in the text. The underlying principle is that the green token number for the correct message candidate is likely to be much larger than the green token number for incorrect candidates.  Despite high accuracy, the heavy computational cost limits the practicality of this approach in real-world applications. This is due to the necessity of enumerating all $2^b$
  potential messages, where $b$ is the bit length of the message.

%faces a significant limitation of high computation cost for extraction.
%The extraction overhead restricts its applicability in real-world scenarios. %, which often require embedding more than 20 bits.

%the hash function generates values distinct from those produced with the correct key, leading to a different green/red list partition for each token. Therefore, the count of green tokens under these circumstances is significantly lower compared to when the correct key is utilized. 

\noindent
\textbf{Limitations of bit assignment-based multi-bit watermark.}   The bit assignment-based multi-bit watermark~\cite{yoo2023advancing} pseudo-randomly assigns each token a \revise{short bit-segment (1-3 bits)} for embedding. Specifically, the segment for the \(i\)-th token is pseudo-randomly determined by the \((i-1)\)-th token. This approach offers high efficiency. However, a significant drawback of this method is its low correctness. The number of tokens used to embed each bit can be quite uneven due to imbalanced token frequencies. As a result, some bit positions may be represented by very few tokens or even none, making it likely to extract incorrect results for those bits.

\revise{Although the imbalance issue of ~\cite{yoo2023advancing} could be mitigated by increasing the segment size, this results in adding bias on a very small vocabulary partition, which significantly degrades text quality.}

\noindent\textbf{Limitations of distortion-free multi-bit watermark.}  
The watermark method in ~\cite{boroujeny2024multi}  achieves the interesting theoretical property of exactly preserving the output distribution of the original LLM. However, our analysis reveals its poor performance when embedding longer messages in practice. This issue arises from its distribution interval encoding strategy. Specifically, the probability of generating the same token for neighboring messages is $(1-2^{1-b})^{17}$, where $b$ is the embeded message length. This probability is nearly 1 for relatively large $b$.  For example, when embedding 20 bits into 200  tokens, the watermarked text will be exactly same for two neighboring messages with a probability greater than 99\%. This makes it almost impossible to distinguish between the two neighboring messages.
%, leading to very low extraction accuracy due to random guessing.  
For more details, please refer to Appendix~\ref{app:analyze_distortion_free}.

\revise{\cite{zamir2024excuse} extended the approach of \cite{christ2023undetectable} to enable distortion-free multi-bit watermarking by additionally feeding the $i$-th bit of the message to the pseudo-random function for the $i$-th token. However, this method is highly vulnerable to  attacks like sentence removal.}

\noindent \textbf{Limitations of block-based multi-bit watermark.}
~\cite{cohenwatermarking} proposed a block-based watermarking method by partitioning generated text into continuous blocks based on entropy. For each block, they embed one bit of the message using existing watermarking schemes such as those in ~\cite{pmlr-v202-kirchenbauer23a}. The major drawback of their solution is its poor robustness against real-world editing. An attacker can simply delete an entire sentence from the generated text, which will likely cause some blocks to be lost, leading to extraction failure. 

\noindent
\textbf{Differences with works adopting ECC.} 
Multiple concurrent works~\cite{fairoze2023publicly,chao2024watermarking,li2024resilient} in zero-bit LLM watermarking also utilize ECC in their constructions. However, their settings and techniques are significantly different from ours. 
%~\cite{fairoze2023publicly,chao2024watermarking,li2024resilient} all focus on the setting of zero-bit watermarking, while ours focuses on multi-bit watermarking. 
~\cite{fairoze2023publicly} achieves the property of public detectability by embedding a cryptographic signature into text. The major difference between \cite{fairoze2023publicly} and our scheme is that they embed each bit using a fixed number of continuous tokens.
%specifically by constraining the hash of tokens with indices ranging from $(i-1)*b+1$ to $i*b$ to be equal to the value of $i$-th bit. 
However, this solution is highly vulnerable to token deletion and insertion, even a single deletion might alter multiple signature bits. 
%Furthermore, the scheme constrains the generated paragraph to contain at least $b \cdot \ell$ tokens when embedding $b$-bits information. Here $\ell$ is the length of signature bits, which typically satisfies $\ell\ge 768$.  This makes it impractical.
~\cite{chao2024watermarking} 
%embeds information based on binarized tokens and viewing token sampling as a binary sampling channel. Their method 
presents theoretical guarantees of high detection probability under bounded edition, while our method further guarantees high probability of perfect message extraction under bounded edition. ~\cite{li2024resilient} adopts BCH code~\cite{bose1960class} in their scheme for AI-generated code watermarking. They embed information by applying code transformations; thus, their method only applies to code LLMs. In contrast, our scheme is applicable to general AI text generation scenarios.

 \subsection{Background on Reed-Solomon codes}

In this subsection, we briefly introduce the background of the error-correction code (ECC) used in our watermarking schemes. Reed-Solomon codes form a large family of error-correction codes that are constructed using polynomials over a finite field. They are widely used in the real-world applications, including digital communication systems \cite{cox2008underwater}, storage devices \cite{ramkumar2021codes}, and cryptography \cite{ames2017ligero}. Reed-Solomon code is optimal in the sense that it achieves the largest possible minimum distance for linear codes of a given size~\cite{singleton1964maximum}. Thus we adopt Reed-Solomon codes~\cite{reed1960polynomial} in this work.

\noindent
\textbf{Definition.}  A Reed-Solomon code is a block code notated as $(n, k, t)_{q^m}$. It takes $k$ symbols from a finite field $\mathsf{GF}(q^m)$ \footnote{$\mathsf{GF}(q^m)$ stands for Galois field (finite field) of order $q^m$.} as the input message and then outputs $n$ symbols belonging to the same field as the encoded message. As long as no more $t$ symbols are changed in the encoded message, it is guaranteed that the $k$ symbols can be fully recovered. 

In this paper, following the convention of applying Reed-Solomon codes~\cite{westall2010introduction,geisel1990tutorial}, we focus on the case where $q=2$, meaning all symbols are integers in $[0,2^m-1]$.
For a detailed exposition of the construction and fundamental principles of Reed-Solomon code, readers may refer to \cite{clark2013error,lin2021fundamentals}.

\section{Problem Formulation}
We present the formal definition of multi-bit watermarks for LLMs.

\subsection{Problem definition}

Suppose we have an LLM that generates text in response to an input prompt. Given an arbitrarily bit string $K$ (called \emph{message}) whose length is $b$, we aim to embed it into a text generated by the LLM. In particular, we aim to design two functions for a multi-bit watermark algorithm, namely \emph{Embedding} and \emph{Extraction}. The embedding function aims to embed the message $K$ into a text generated by the LLM. The extraction function aims to extract the message from a given text. Formally, we have the following definition.

\begin{definition}[Multi-bit Watermarking]
Given an LLM, a message $K$, and a prompt token sequence $\textbf{P} = [P_{0}\cdots P_{N-1}]$,  a multi-bit watermark algorithm consists of the following two functions:
\begin{align}
    \textbf{S} &= \text{Embedding}(\mathsf{LLM},K,\textbf{P}) \\
    K' &= \text{Extraction}(\textbf{S}).
\end{align}
where $\textbf{S} =[S_{0},\cdots, S_{T-1}]$ is the generated token sequence with length $T$.
\end{definition}

\subsection{Design goals}
We aim to design a multi-bit watermark algorithm to achieve the following goals.  

\noindent
\textbf{Correctness.} This goal means the algorithm could accurately extract a message embedded in a watermarked text. Formally, this requires the following conditional probability to be equal (or close) to 1.
\begin{align}
&\Pr(\text{Extraction}(\textbf{S}) = K |\textbf{S}= \text{Embedding}(\mathsf{LLM},K,\textbf{P})). %\approx 1.
%\max \Pr(K=K')
\end{align}

%\jiaheng{We should formulate Correctness and Robustness using the notation of Embedding and Extraction. }

\noindent
\textbf{Robustness.}  
%\begin{definition}
%Define $\operatorname{D}(y, z)$ as the \textbf{edit distance} between sequence $y$ and sequence $z$. $\operatorname{D}(y, z)$ quantifies the number of basic operations required to transform a sequence $y$ into another sequence $z$. These operations include "insertion", "deletion", and "replacement" of tokens.
%\end{definition}
This goal means the algorithm is resilient against post-processing operations on watermarked text,  such as word addition, word deletion, synonym substitution, and paraphrasing. Let $\textbf{S}' $  represent the edited version of the original text $\textbf{S}$. The edit distance between $\textbf{S}$ and $\textbf{S}'$ is bounded by $\mathsf{D}(\textbf{S}, \textbf{S}') \leq \eta$, which means $\textbf{S}$ can be transformed into $\textbf{S}'$ within $\eta$ basic operations ("insertion", "deletion", and "replacement" of tokens). Then the robustness goal requires the following conditional probability is equal (or close) to 1. 

\begin{small}
    \begin{align}
&\Pr(\text{Extraction}(\textbf{S}') = K |\textbf{S}= \text{Embedding}(\mathsf{LLM},K, \textbf{P}), D(\textbf{S},\textbf{S}')\le \eta)
%\approx 1.
%\max \Pr(K=K')
\end{align}
\end{small}

%Formally, robustness (parameterized by $\eta,\alpha$) is defined as $\Pr(\mathsf{Extraction}(S')=K)\ge 1-\alpha$.

%Let $S'$ be the editted version of $S$, the edition on $S$ is bounded, namely $\mathsf{D}(S,S')\le \eta$. The algorithm maximizes the probability of $S'$  accurate extraction $\Pr(\mathsf{Extraction}(S')=K)$. \\

%Let $S'$ represent the text obtained by an attacker through post-processing $S$. The algorithm seeks to maximize the probability of correct extraction of information, denoted as $Pr(\text{extract}(S') = K)$.}

\noindent
\textbf{Efficiency.} The efficiency goal means the algorithm can embed/extract a message into/from a generated text efficiently. 
%, and it can extract the bit string from a given text in a short time.

%We briefly review the state-of-the-art watermarking technique used in \cite{pmlr-v202-kirchenbauer23a}, then discuss the limitations of existing multi-bit watermarking, and finally introduce the basics of BCH error-correction code~\cite{bose1960class}.

%In this work, we mainly consider the embedded information with length $\le 32$ without loss of generality. Note that our method is applicable even if $> 32$, in that case we can use BCH schemes like $[63,39,4], [63,51,2]$ to embed messages. 

\section{Methodology}

In this section, we will discuss the design insights and algorithmic details of our watermarking method, along with the high-level concepts behind its provable robustness.

\begin{table*}[!t]
\centering

\caption{Example of our watermarking and extraction results. The content in the "Prompt" column comes from datasets. The "Real Completion" column represents the ground truth completion from datasets. The "Watermarked Text" column represents text embedded with 20-bit information by our watermarking algorithm. %The "$\mathsf{COUNT}$" column contains the $\mathsf{COUNT}$ matrix obtained during watermark extraction, grids corresponding to correct information are in green, grids corresponding to wrong information are in red. 
 }
%\vspace{-1em}
\resizebox{1.0\hsize}{!}{
\label{tab:watermark_extract_example}
\begin{tabular}{c|c|c|c|c}
\hline
Prompt    & Real Completion & Watermarked Text & \makecell[c]{ Original\\ bits}  &  \makecell[c]{Extracted \\ bits} \\ \hline
\makecell[l]{
... The bin floor is where the grain \\
is stored before being ground into \\
flour. The sack hoist mechanism is \\
housed on this floor, driven from \\
the wooden crown wheel on the \\
floor below by belt.
}
&
\makecell[l]{
The stone floor is where the two pairs \\
of underdrift French Burr millstones are \\
located. The stones are 4 feet 4 inches \\
(1.32 m) and 4 feet 7 inches (1.40 m) \\
diameter. The wooden crown wheel is \\
located towards the ceiling. [...continues]
}
&
\makecell[l]{
The floor beneath that is the grinding \\
floor, containing pairs of millstones \\
capable of producing bushels of flour \\
per day. This floor is reached by a \\
spiral stairs from within the dust floor.\\
The crown wheel is on this level [...continues]
}
&
{00110100110100111010} & {00110100110100111010} \\ \hline
\end{tabular}
}

\end{table*}

%\vspace{-1em}

\subsection{Design insights} 
\label{sec::insight}

%In the workflow of our watermarking method, users query the LLM service with prompt to generate text. The service provider assigns each user a unique user ID and intend to embed information such as user ID and the current date in the generated text. Later when the service provider notices suspicious LLM-generated text online used for malicious purpose, they could employ our watermark extraction function to identify the LLM generated text and further trace back to the malicious user who generated the text.

\subsubsection{Insights from previous works}

%The problem of multi-bit watermarking can be viewed as how to effectively use the redundancy in selecting each token to embed the message.
%To gain insights for designing a multi-bit watermarking scheme that meets our four design goals, we first analyze the weaknesses of concurrent work by Yoo et al.~\cite{yoo2023advancing}. In this work, during generation, each token is used to embed one bit of the message and the bit position each token should embed is determined by the hash of its previous token. This approach naturally partitions the $T$ tokens into $b$ groups, with each group embedding a specific bit. 
To further gain insights into designing our multi-bit watermarking scheme, we dive into the design of several previous works~\cite{fairoze2023publicly,cohenwatermarking,yoo2023advancing}. %could be viewed as special cases of this framework.  
To embed a $b$ bit message, these works all partition the message into $b$ individual bits, and partition the tokens into $b$ groups to embed each bit respectively. Specifically, ~\cite{fairoze2023publicly,cohenwatermarking} partitions continuous tokens into groups. They use each group of continuous tokens to embed each individual bit. However, the major drawback of continuous token partition is weak robustness. Because removing an entire sentence may cause the lost of a whole group of tokens, leading to the extraction error of the corresponding bit. 
 
 The algorithm in~\cite{yoo2023advancing} is different in that \revise{under typical setting, it assigns each token a 2 bit segment of the message to embed based on the hash of the previous token}. In this way, tokens are divided in a pseudo-random manner. Due to the pseudo-randomness in the bit segment embedded by each token, this method achieves better robustness against attacks such as sentence removal. However, this method fails to achieve high accuracy when embedding longer messages. Deeper inspection into the bits extracted incorrectly reveals that these bits are often allocated few tokens or even none at all. 
This imbalance in token allocation significantly hinders \cite{yoo2023advancing} to achieve high correctness, because when a \revise{segment} is embedded using only a few tokens, there is a considerable probability of extracting the  \revise{segment} incorrectly.
%Token imbalance in bit assignment harms correctness because when a bit is embedded using only a few tokens, there is a considerable probability of extracting the bit incorrectly. 
In the extreme case,  when no tokens are assigned to a \revise{segment}, the \revise{segment}’s value  can only be guessed randomly.

Further investigations reveal that this imbalance is rooted in the imbalance in natural language token frequency~\cite{wolleb2023assessing,gu2020token}. Because the distribution of each token's previous token is severely imbalanced, the bit positions that frequent tokens map to are assigned more tokens. Assume we want to embed 32 bits into the generated text. Ideally, each bit is expected to be selected with a probability of $1/32$. However, frequent tokens like "the" already exceed this balanced probability of $1/32$ (the frequency of "the" is 3.4\% as measured on the C4~\cite{raffel2020exploring} dataset). Other less frequent tokens may still be mapped to the same bit as "the" with nearly 1/32 probability. This causes the probability of the bit position corresponding to $\mathsf{hash}(``the")$ being chosen to be at least twice as high as in the uniform case. Consequently, some bit positions are rarely
selected and therefore fewer tokens are used to embed these bits.

However, we also discover that this problem is not that severe when embedding shorter messages. For example, if we only need to pseudo-randomly assign each token to one of the 5 group, the hash result of a popular token with 1\% frequency only has a relatively small influence on the balance of token allocation. This raises the question: how can we reduce the number of groups into which we divide tokens while still embedding the same message?

%Using fewer tokens to embed one bit will increase the probability of extracting wrong value for that bit, because the wrong candidate is more likely to own equal or even more green tokens than the correct bit value, given fewer total tokens. For example, when only 1 token is used to embed a bit's value, the wrong bit value also has $\frac{1}{2}$ probability to own 1 green token. Therefore, under half of the cases, we can't distinguish whether 0 or 1 is the correct bit value. In the extreme case when no token is assigned, we can only randomly guess the value of this bit.

%, which is the max green tokens number a candidate can achieve in this case. 

%The drawback of embedding long bit-strings partially cancels out the benefit of ECC on correcting errors. This makes the above approach far away from being an ultimate solution to this problem.

\subsubsection{Key ideas of our watermark design}

Our solution is to pack multiple bits together into \textbf{bit segments}. \revise{The difference between our approach and \cite{yoo2023advancing} lies in the method of embedding a segment into a token. \cite{yoo2023advancing} embeds a $d-$bit segment by adding bias to a $\frac{1}{2^d}$ fraction of vocabulary. Increasing the segment size of~\cite{yoo2023advancing}  (e.g., size 8) could mitigate the assignment imbalance but will split the vocabulary into 256 parts per step. This approach severely restricts word choice for each token, leading to a decline in text quality. Due to this limitation, rather than using longer bit segments, \cite{yoo2023advancing} chose to assign a 2-bit segment to each token. Contrarily, our solution embeds the assigned multi-bit segment into a token by incorporating it as an additional input to the hash function.}  Our approach effectively reduces the number of segments compared with \cite{yoo2023advancing}, while still embedding the same message.
Note that each token now simultaneously embeds every bit in its assigned segment. Consequently, the number of tokens embedding each bit is much more balanced.

%For example, when there are 32 token groups, a token appearing with $1\%$ frequency has quite considerable influence to the balance of token numbers in each group. However, when there are only 4 token groups, the influence of this token will be much smaller. 
%To reduce the number of slots while still embedding the same number of bits, we propose to pseudo-randomly assign tokens to \textbf{bit segments} instead of to \textbf{invidivual bits}. 

We offer an toy example to illustrate how the balance is improved. Assume that we intend to embed a 16-bit message into 200 tokens. The number of tokens pseudo-randomly allocated for each bit is $[4,20,16,8,14,14,6,22,16,10,8,14,15,3,7,23]$. However, if each continuous 4 bits are packed together to form segments, the number of tokens allocated to each segment is $[48,56,48,48]$. 
Take the first segment as an example. Since the 48 tokens allocated to it simultaneously embed the first segment of 4 bits, each bit in the first segment is conceptually embedded by $48/4=12$ tokens.
Now the number of tokens embedding each bit becomes $[12,12,12,12,14,14,14,14,12,12,12,12,12,12,12,12]$. From this example, we can obtain that segment-based embedding can significantly improve the balance in the distribution of tokens embedding each bit.

We denote the message we intend to embed as $K$, and its bit length as $b$. The watermarked text generated by the LLM comprises $T$ tokens. 
Formally, our method divides the original message $K$ into $k$ segments. Without loss of generality, we assume $b$ is a multiple of $k$. For each token, we pseudo-randomly assign the segment it embeds based on its previous token.  This approach partitions the $T$ tokens into $k$ groups instead of $b$ groups. Each group of tokens is assigned a segment with a value in the range $[0,2^{\frac{b}{k}}-1]$ to embed.

Next, we need to embed each segment into its allocated token group. 
Inspired by ~\cite{fernandez2023three,wang2023towards}, we can use the segment's value as an additional input 
besides the previous token to the hash function to compute a random seed $s$. Seed $s$ is then used to sample the current token's green token list. During extraction, for each segment, we enumerate the possible values in $[0,2^{\frac{b}{k}}-1]$ and choose the value that maximizes the count of green tokens as the segment's extracted value. 

The underlying principle is that the number of green tokens associated with the correct value is likely to be much larger than that of other values. 
The advantage of correct value comes from the significant differences among the green lists pseudo-randomly generated with different seeds. Through biasing the tokens in green list, each token embedding this segment is highly likely to belong to the green list of the segment's value. On the other hand, for every other value, the probability for each generated token to belong to this value's green list is only $\frac{1}{2}$. This probability difference allows us to distinguish between the correct value and other values.

%the segment value extracted in this way is likely to be the correct value of the segment.
\begin{figure}[h]

\centering
\subfloat{\includegraphics[width=0.48\textwidth]{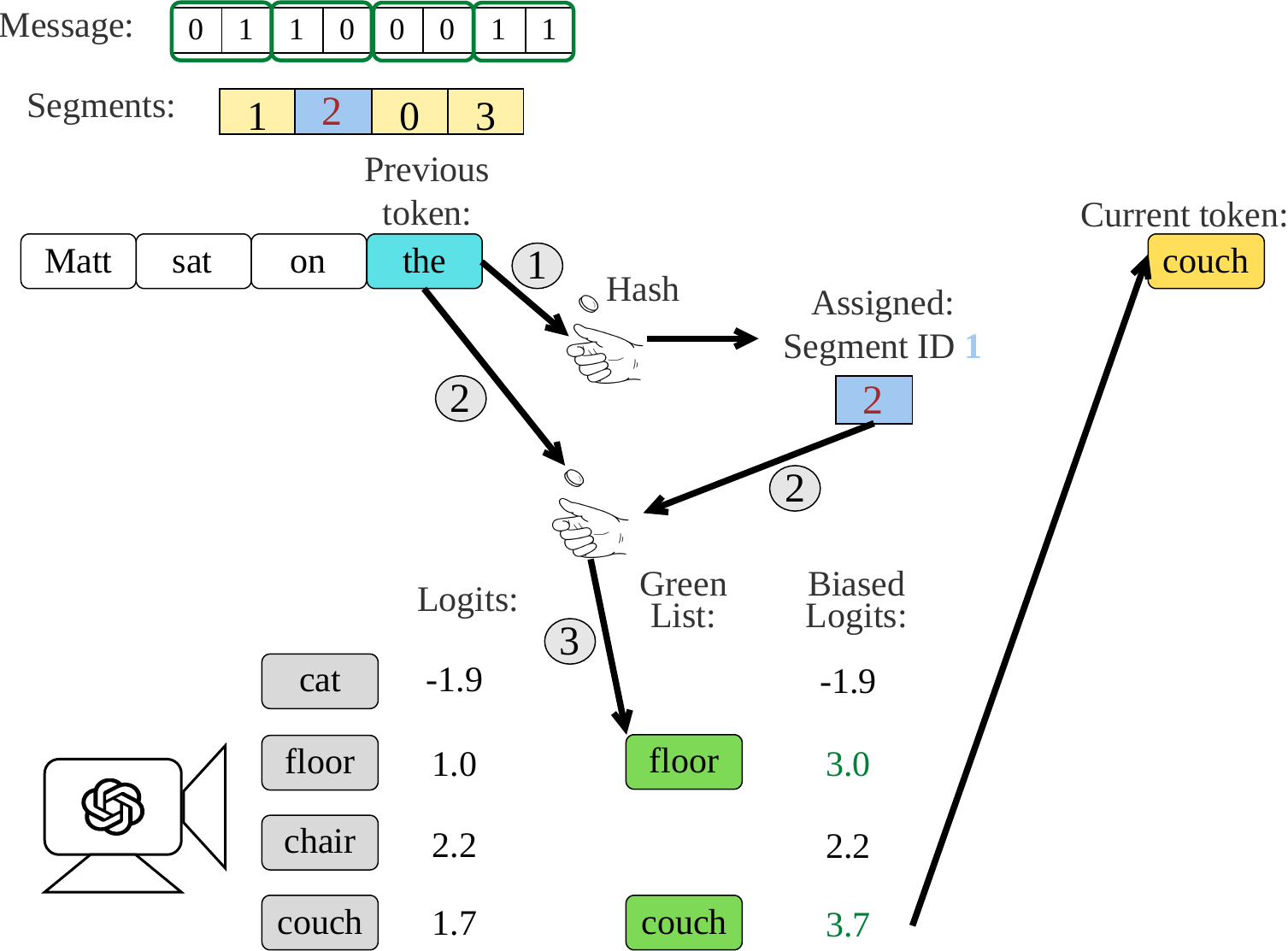}}

\caption{Simplified example of pseudo-random segment assignment-based watermark embedding. \protect\scalerel*{\textcircled{\raisebox{-0.7pt}{1}}}{1} Determine the index of segment to embed based on previous token. \protect\scalerel*{\textcircled{\raisebox{-0.7pt}{2}}}{2} Obtain seed $s$ based on previous token and segment value. \protect\scalerel*{\textcircled{\raisebox{-0.9pt}{3}}}{3} Select green list using seed $s$.}
\label{fig:embed_example}
\vspace{-1em}
\end{figure}

We present a sketch of our segment assignment-based watermark embedding function in Figure \ref{fig:embed_example}. Suppose the message to embed is "01100011" and we divide the message into 4 segments with values $[1, 2, 0, 3]$. For simplicity, we only show 4 tokens in the LLM vocabulary. After the LLM is provided with prompt "Matt sat on the" to generate the next token, it outputs the logits of these 4 tokens as $[-1.9, 1.0, 2.2, 1.7]$. To embed the watermark, the previous token "the" will first be fed into a hash function to determine the index of segment that current token should embed.
In this case the segment index is 1, and therefore, the segment value we embed in the current token is 2. In the second step, we obtain the seed $s$ using hash function taking the previous token "the" and the segment value 2 as input. This seed is used to select the green token list from the vocabulary. In our example, the green list contains 2 tokens "floor" and "couch". We add bias $\delta = 2$ to all green tokens, resulting in the modified logits vector $[-1.9, 3.0, 2.2, 3.7]$. Finally, we use greedy sampling to choose "couch" as the next token. Through biasing tokens in green list, instead of token "chair", "couch" is selected as the output token.

 Our proposed approach is much more efficient during extraction  compared with ~\cite{wang2023towards,fernandez2023three}. ~\cite{wang2023towards,fernandez2023three} have extraction complexity $O(2^b)$ due to the need to enumerate over the whole potential message space. In our approach, the extraction complexity is only $O(k\cdot 2^{\frac{b}{k}})$, as the enumeration is conducted for each $\frac{b}{k}$-bits segment individually. \revise{Our method provides a systematic way to trade-off between watermark accuracy and computational efficiency by using different segment number $k$. A smaller $k$ improves watermark accuracy by allowing more tokens to embed each segment but increases computational cost. Conversely, a larger $k$ reduces extraction complexity but also slightly sacrifices the watermark accuracy. %Our scheme also unifies both message enumeration-based methods~\cite{wang2023towards, fernandez2023three} and bit assignment-based methods~\cite{yoo2023advancing}, treating the former as a special case where $k = 1$ and the latter as a special case where $k = b$.
 }

\subsubsection{Further improvements} 
Although our algorithm has shown significant advantages over existing ones, we aim to enhance its real-world applicability by maximizing its correctness and robustness without compromising extraction efficiency. To achieve this, we have developed two additional techniques.

Our first improvement aims to further eliminate the segment assignment imbalance induced by the imbalance of different token's frequency. 
We design a dynamic programming-based algorithm to compute a frequency-balanced mapping to assign tokens to different segments based on natural token frequency.  During embedding and extraction, we leverage this mapping to assign each token a segment to embed. After taking tokens' natural frequency into consideration, we further alleviate the imbalance in segment assignment.

Additionally, our algorithm still faces a challenge: when the attacker's modifications are substantial enough to distort the value embedded in a segment, the embedded message may no longer be correctly extracted. To address this issue, our second improvement integrates Reed-Solomon code~\cite{reed1960polynomial} into our watermarking scheme.  We opt for Reed-Solomon code in light of its effectiveness in correcting burst errors. This property naturally aligns with the needs of our watermarking scheme in that when extraction error occurs in our scheme, they often take the form of burst errors, where only one or two segments are incorrect, but multiple bits within those segments may be wrong.

Based on our segment partition, the input message can be viewed as a sequence of $k$ segments (each of which is an integer in $[0,2^\frac{b}{k}-1]$). We use Reed-Solomon code scheme to encode these $k$ segments into $n$ segments. Instead of embedding the original $k$ segments, we embed these $n$ segments into the generated text. The property of Reed-Solomon code ensures that if no more than $t = \lfloor\frac{n-k}{2} \rfloor$ segments are extracted incorrectly from the text, the final message decoded from the $n$ segments will be guaranteed correct. Adopting Reed-Solomon code simultaneously improves the correctness and robustness of our method.

We illustrate the concrete technique of our multi-bit watermarking in the following section. We present the high-level workflow of our multi-bit watermarking algorithm in Figure~\ref{fig:concept}.

\subsection{Design details of multi-bit watermarking}

We first illustrate the details of balanced segment assignment used in our watermark embedding and extraction function. Then we discuss the procedure of embedding a message $K$ with $b$ bits into a text generated by an LLM. Finally, we introduce how to extract the message from given text.

\noindent
\textbf{Balanced segment assignment.} Balanced segment assignment can be transformed into the following problem: given the frequency $p_i$ of token $i$ appearing in natural text, we need to pseudo-randomly map each $i \in [0,\cdots, |V|-1]$ to a segment $j \in [0, \cdots, \hat{n}-1]$, and make the probability of each segment to be chosen as close to uniform as possible.

A strawman way to reformulate this problem is to divide $p_0,\cdots,p_{|V|-1}$ into $\hat{n}$  subsets and minimize the variance of the sum of each subset. 
%[The drawback of this formulation is that
%are: (1) the search space of possible solutions to this problem is $\hat{n}!$, which may not be big enough when $\hat{n} \leq 10$ and introduce security concerns (brute force enumeration over the mapping table). (2) this problem 
%this problem belongs to the partition problem~\cite{hayes2002computing,mertens2003easiest}, which is NP-hard, indicating the difficulty of finding the optimal solution.]
However, such a problem can be reduced to the partition problem~\cite{hayes2002computing,mertens2003easiest}, which is NP-hard. 

%[We alternatively solve a slightly modified version of this problem.]\zhihua{
To that end, we propose an alternative version of the problem. Specifically, assume that the IDs of $|V|$ tokens are pseudo-randomly shuffled using private key $\mathsf{sk}$, ranging in $\{0,\cdots,|V|-1\}$; we partition the vocabulary $V$ into $\hat{n}$ groups, with tokens in each group having continuous IDs. We denote the frequency of token with ID $i$ after shuffle as $p'_{i}$. Our target is to minimize the variance of the sum within each group, thereby ensuring the plan to achieve optimal balance.

Through derivation, the goal is equivalent to minimizing the sum of the square of each group's sum. Denote the sum of each group $i$ as $\mathsf{Sum}_i$. We aim to minimize $\sum_{i=0}^{\hat{n}-1} \mathsf{Sum}_i^2$.

We propose to solve this problem by dynamic programming. For the tokens with ID in $[0, i]$ divided into $j$ groups, denote $D(i,j)$ as the minimal squared sum of these groups' internal sum. Denote the starting token ID of the $j-$th group as $\ell$. The first $\ell$ tokens are partitioned into $j-1$ groups. $D(i,j)$ must transition from some $D(\ell-1,j-1)$. Thus, we have:
%\begin{align}
    $D(i,j)=\min_{\ell=j-1,\cdots, i} D(\ell-1, j-1)+ (\sum_{u=\ell}^ i p'_{u})^2$
%\end{align}

The computation of each $D(i,j)$ requires $O(|V|)$ complexity. Thus, the complexity of computing $D(|V|-1, \hat{n})$ will eventually become $O(\hat{n}\cdot |V|^2)$. Since this dynamic programming only needs to be done once for each Reed-Solomon code scheme, the computation overhead is acceptable. It can be further speeded up by leveraging parallel computing.

\noindent
\textbf{Design of \emph{Embedding} function.} Our Embedding function consists of the following steps.

\emph{Step I:}
In the first step, we aim to select an ECC scheme that has enough payload and satisfies the minimal code rate $R_c$ and minimal recover rate $R_r$. We first filter for all schemes $ (n,k,t)_{2^m} \in \mathcal{F} $ satisfying $km=b$, $\frac{k}{n} \geq R_c$ and $\frac{t}{n}\geq R_r$. Among these candidate schemes, 
%we further select the code schemes that has sufficient payload length ($k \cdot m \geq b$) and recover rate ($\frac{t}{n} \geq R_r$). Finally, 
we choose the code $ (n,k,t) $ with the lowest $n$ to embed. %among them. 
%This strategy also maximizes the code rate of our embedding. 
For simplicity, we use $(\hat{n}, \hat{k}, \hat{t})_{2^{\hat{m}}}$ to denote the selected code scheme. Algorithm \ref{alg:choose_rs} in Appendix \ref{sec:app:pseudo_code} presents the complete ECC scheme selection process.

After we determine the code scheme to use, we compute the pseudo-random token-to-segment mapping $M: [0, |V|-1] \rightarrow [0, \hat{n} - 1]$ with a secret key $sk$ using the above-mentioned balanced segment assignment algorithm. Secret key $sk$ is used to pseudo-randomly shuffle the tokens, therefore protecting our mapping from being discovered by the attacker.

%Suppose we are given a minimum code rate $R$, which measures the proportion of the embedded bit string that is non-redundant. 

%In this study, we conduct an exhaustive enumeration of all 

%Initially, our consideration is limited to codes $ f = (n,k,t) \in \mathcal{F} $ for which $ k \geq b $, ensuring a sufficient information payload. Additionally, we impose a minimum code rate $ R $, defined by the ratio $ \frac{b}{n} $. This necessitates the inclusion of the constraint $ \frac{b}{n} \geq R $ to preclude excessive redundancy of bits. In cases where multiple codes still qualify, our selection criterion favors the code $ f $ with the highest recovery rate $ \frac{t}{n} $, thereby opting for the code with the most robust error correction capability.
%The information length may not exactly match the Reed-Solomon code. If $b\ne \hat{k}\cdot \hat{m}$, we append "0" bits to $K$ to fill up length $\hat{k}\cdot \hat{m} $. W

\emph{Step II:}
In the second step, we use the previously selected Reed-Solomon code scheme $(\hat{n}, \hat{k}, \hat{t})_{2^{\hat{m}}}$ to encode $K$, resulting in an ECC-encoded sequence $E$ (referred to as the \emph{encoded message}) with a length of $\hat{n}$, where each element is within the range $[0, 2^{\hat{m}}-1]$.

\emph{Step III:}
In the last step, we embed the encoded message $E$ into text generated by an LLM. The process of embedding $E$ into a text is as follows: at the $i$-th step of token generation, we query the token-to-segment mapping $M$ with previous generated token $S_{i-1}$ to obtain the segment index $p$. We denote the $p$-th segment as $E[p]$. Then we compute a random seed $s=\mathsf{hash}(sk,  S_{i-1}, E[p])$ with secret key $sk$, the previous token $S_{i-1}$ and $E[p]$. Secret key $sk$ is used to strengthen the security of our scheme, similar to~\cite{pmlr-v202-kirchenbauer23a}. Without knowing the secret key, a potential attacker can only randomly guess whether a token belongs to the previous tokens' green list or red list. We utilize seed $s$ to pseudo-randomly select the green list $G$ with length $\frac{|V|}{2}$ from vocabulary $V$.

At the $i$-th step, after the LLM outputs the logits vector $v_i$, we add bias term $\delta$ to the logits of all tokens in $G$.  Then we use the LLM decoding algorithm to sample the output token $S_i$ from modified logits $\hat{v}_i$. The entire embedding procedure of our watermarking method is described in Algorithm~\ref{alg:encode}.  

%, major differences of our embedding procedure and Kirchenbauer et al~\cite{pmlr-v202-kirchenbauer23a} are highlighted in blue.

%At step $i$, if $B[p]$ is 1, $G$ is set as $V_1$; otherwise, $G$ is set as $V_0$. In this manner, the green list $G$ is determined by both pseudo-random seed $s$ and encoded message bit $B[p]$. 

\noindent
\textbf{Design of \emph{Extraction} function.}  
Let $E'$ denote the segment sequence directly extracted from the token string $S$. During extraction, we first initialize the matrix $\mathsf{COUNT}$ counting green tokens for each potential value of each segment as zero. For the $i$-th token, we first obtain its corresponding segment's index $p$ through querying the token-to-segment mapping $M$ using the previous token. Then for each possible value of $p$-th segment $j \in [0, 2^{\hat{m}}-1]$, we compute its corresponding seed $s$ with secret key $sk$, previous token $S_{i-1}$ and value $j$. Seed $s$ is subsequently used for selecting the green list $G$. If $i$-th token $S_i$ belongs to the green list $G$ of value $j$, we add 1 to $\mathsf{COUNT}[p][j]$. After we enumerate all the tokens, for each position $p$, we determine the value of $E'[p]$ by finding $j$ which maximizes $\mathsf{COUNT}[p][j]$.

%For value $j$, we compute its corresponding version of $G$.  For the set of $j$ which $S_i$ belongs to $G$, we add one to $\mathsf{COUNT}[p][j]$ for $B'_p=j$. After we enumerate the string tokens, for each position $p$, we determine the value of $B'[p]$ by finding $j$ which maximizes $\mathsf{COUNT}[p][j]$. 

%is 0 or 1 by counting the votes, namely checking whether $\mathsf{COUNT}[p][0]$ is greater than $\mathsf{COUNT}[p][1]$, or vice versa. When they are equal, we randomly break the tie.

%Each time we compute $G$ based on if $S_i \in G$ then it "votes" to $B'[p]=0$ by adding 1 to $\mathsf{COUNT}[p][0]$, otherwise it "votes" to $B'[p]=1$ by adding 1 to $\mathsf{COUNT}[p][1]$. 
%and we count the contributions to all message positions in $\{0,1,\cdots,\hat{n}-1\}$ and record in the $\mathsf{COUNT}$ matrix. 
%After we enumerate the string tokens, for each position $p$, we determine whether $B'[p]$ is 0 or 1 by counting the votes, namely checking whether $\mathsf{COUNT}[p][0]$ is greater than $\mathsf{COUNT}[p][1]$, or vice versa. When they are equal, we randomly break the tie. 

After we decided $E'$, we use $(\hat{n},\hat{k},\hat{t})_{2^{\hat{m}}}$ to decode $E'$ to obtain $K'$. 
The entire extraction procedure of our watermarking method is described in Algorithm ~\ref{alg:decode}.

%; major differences between our extraction procedure and Kirchenbauer et al. ~\cite{pmlr-v202-kirchenbauer23a} are highlighted in blue.

Table~\ref{tab:watermark_extract_example} illustrates an example of our watermark pipeline. The example show that the embedded message can be accurately extracted from the watermarked text, while the watermarked text itself maintains good quality.

\begin{algorithm}[tb] \small
%\jiaheng{Remove the blue color. }   
\caption{\emph{Embed multi-bit watermark}}
   \label{alg:encode}
\begin{algorithmic}
   \STATE {\bfseries Input:} Autoregressive language model LLM; generate length $T$; prompt sequence $P_{0}\cdots P_{N-1}$; vocabulary set $V$; bias $\delta$; set of Reed-Solomon code $\mathcal{F}$; message $K$ with $b$ bits ; minimal code rate $R_c$; minimal recover rate $R_r$; a hash function $\mathsf{hash}$; secret key $sk$\\
   \STATE {\bfseries Output:} Generated sequence $S_{0}\cdots S_{T-1}$
   \\
   \STATE Initialize $\textbf{S}=$""\\
   \STATE Select optimal $(\hat{n}, \hat{k}, \hat{t})_{2^{\hat{m}}}$ using Algorithm~\ref{alg:choose_rs} under $R_r$ and $R_c$
    %\STATE Extend $K$ to $K'$ with length $\hat{k}\cdot \hat{m}$, by appending "$0$"s \\
    \STATE Obtain ECC-encoded segment sequence: $E=(\hat{n}, \hat{k}, \hat{t})_{2^{\hat{m}}}.\mathsf{encode}(K)$\\
    \STATE Calculate pseudo-random token-to-segment mapping $M: [0,|V|-1] \rightarrow [0,\hat{n}-1]$ with $sk$\\
    \FOR{$i=0,1,\cdots, T-1$} 
    \STATE Compute logits, $v_i = \mathsf{LLM}(P_{0}\cdots P_{N-1}, S_{0}\cdots S_{i-1})$ \\
    \STATE  Compute segment index $p=M[S_{i-1}]$ \\
    %by querying token-to-segment mapping $M$ with %$S_i$\\
    %\STATE  Compute random seed $s=\mathsf{hash}(sk, S_{i-1}, E[p]) $\\
    %\STATE Take the first $\hat{m}-1$ bits of $E[p]: e=\lfloor\frac{E[p]}{2}\rfloor$\\
    \STATE Compute random seed $s=\mathsf{hash}(sk, S_{i-1}, E[p]) $\\
    \STATE Use seed $s$ to select green list $G \subset V$, where $|G|=\frac{|V|}{2}$ \\
    %\STATE Determine green list:\\
    %\STATE Set green list $G$ such that $G_{i}[V_0] = 0$ and $G_{i}[V_1] = 1$
    % \STATE Let $G'_{i}$ be a tensor such that $G'_{i}[V_0] = 0$ and  $G'_{i}[V_1] = 1$
    % \STATE Set green list $G \leftarrow \mathsf{roll}(G', -E[p])$\\
    \STATE Bias the logits: \\
    \FOR{$w = 0,1,\cdots,|V|-1$} 
    \IF{$w\in G$}
    \STATE $\hat{v}_i[w]=v_i[w]+\delta$\\
    \ENDIF 
   % \STATE \textcolor{red} {$\hat{v}_i[w]=v_i[w]+\delta \vmathbb{1} (w\in G)\oplus(E[p] mod 2)$}\\
    %\STATE where  $ \vmathbb{1}(\cdot) 
    \ENDFOR
    \STATE Append $S_{i}$ to $\textbf{S}$ by sampling from logits $\hat{v}_i$ \\

    \ENDFOR
    \STATE \textbf{return} $\textbf{S}$\\
\end{algorithmic}
\end{algorithm}

\begin{algorithm}[tb] \small
   \caption{\emph{Extract multi-bit watermark}}
   \label{alg:decode}
\begin{algorithmic}

   \STATE {\bfseries Input:} Text length $T$; text sequence $S_{0}\cdots S_{T-1}$; vocabulary set $V$; Reed-Solomon code scheme $(\hat{n},\hat{k},\hat{t})_{2^{\hat{m}}}$; pseudo-random token-to-segment mapping $M: [0,|V|-1] \rightarrow [0,\hat{n}-1]$; hash function $\mathsf{hash}$; secret key $sk$; \\
   \STATE {\bfseries Output:} {Extracted message $K'$ with $b$ bits}\\
    \STATE Initialize $\mathsf{COUNT}$ as $\hat{n} \times 2^{\hat{m}}$ matrix with 0 \\
    %\STATE Define $E'$ as $\mathsf{GF}(2^{\hat{m}})$ vector with length $\hat{n}$\\
    \STATE Define $E'$ as integer vector with length $\hat{n}$\\
    \FOR{$i=0, 1,\cdots, T-1$} 
    \STATE  Compute segment index $p = M[S_{i-1}]$\\
    \FOR{$j = 0, 1, \cdots, 2^{\hat{m}}-1$}
        \STATE Compute random seed $s=\mathsf{hash}(sk, S_{i-1}, j)$
        \STATE Use seed $s$ to select set $G \subset V$, where $|G| = \frac{|V|}{2}$
        \IF{$S_i\in G$}
        \STATE {$\mathsf{COUNT}[p][j] \mathrel{+}=  1$} \\
        \ENDIF
    \ENDFOR \\
    % \STATE Let $G'_{i}$ be a tensor such that $G'_{i}[V_0] = 0$ and  $G'_{i}[V_1] = 1$
    % \STATE $\mathsf{COUNT}[p] = \mathsf{COUNT}[p] + \mathsf{roll}(G'_{i}, -S_i)$
    
    \ENDFOR \\
    \FOR {$p=0,1,\cdots,\hat{n}-1$ }
    \STATE $E'[p] \leftarrow \arg\max_{j \in [0, 2^{\hat{m}}-1]}  \mathsf{COUNT}[p][j]$ 
    \ENDFOR 
    \STATE $K'=(\hat{n}, \hat{k}, \hat{t})_{2^{\hat{m}}}.\mathsf{decode}(E')$  %\\
    \STATE \textbf{return} $K'$\\
\end{algorithmic}

\end{algorithm}

\subsection{Theoretical robustness analysis}
\label{sec:theory_robust}
In this section, we demonstrate the robustness of our watermarking scheme against editing attempts (insertion, deletion, or substitution of a token). This represents the first LLM multi-bit watermarking scheme that provides a non-trivial provable robustness bound under \textbf{edit distance}~\cite{navarro2001guided} for every generated sentence.  Before presenting the formal theorem, it is essential to first introduce one notation.

\begin{definition}
Define $p_i^{(k)}(x,y,\textbf{c},\textbf{d})$ as the probability that exactly $i$ segments in the first $k$ segments are extracted incorrectly. The condition is  $x$ tokens are added to the groups of the first $k$ segments, and $y$ tokens are deleted from them, initial allocated token numbers for each segment is $\textbf{c}=(c_1,\cdots,c_n)$ and initial green token numbers for each segment is $\textbf{d}=(d_1,\cdots,d_n)$. 
\end{definition}

%We present the main result in Theorem~\ref{thm:robust_theorem}. 

We provide our main theorem as follows:
\begin{theorem}
\label{thm:robust_theorem}

For text paragraph $S$ generated by Algorithm \ref{alg:encode} before any editing. Denote embedding information $K\in \{0,1\}^{km}$ with token number $T$, error-correction code $(n, k, t)_{2^m}$, allocated token number for each segment $(c_1,\cdots,c_n)$ and green token number for each segment $(d_1,\cdots,d_n)$. Attacker modifies $S$ within edit distance budget $\eta$ to obtain $S'$ such that $D(S, S') \le \eta$.  We have 
    \begin{align}
\label{formula:robust_prob_RHS}
\begin{aligned}
    & \Pr(\mathsf{Extraction}(S')\neq K) \le 1-\sum_{i=0}^t p_i^{(n)}(2\eta,2\eta,\textbf{c},\textbf{d})  
\end{aligned}
\end{align}

\end{theorem}

Through the theorem, we demonstrate that when the attacker modifies the watermarked text within an edit distance budget, with quite a small probability, the extracted information will be manipulated. Due to the space limitation, the proof of the theorem is available in Appendix~\ref{appendix:proof}.

We define the \textbf{robust bound} for a generated paragraph with allocated token number for each segment $\textbf{c}=(c_1,\cdots,c_n)$ and green token number for each segment $\textbf{d}=(d_1,\cdots,d_n)$ as follows:
\begin{definition}
\label{define:robust}
    The robust bound $B$ under error rate $\alpha$ is the largest $\eta$ that Theorem ~\ref{thm:robust_theorem} can guarantee the extraction error rate does not exceed $\alpha$. Formally, we have: 
    \begin{align}
        \begin{aligned}
            B=\max\{\eta | 1-\sum_{i=0}^t p_i^{(n)}(2\eta,2\eta,\textbf{c},\textbf{d})\le \alpha \}
        \end{aligned}
    \end{align}
\end{definition}
The robust bound $B$ can be determined by performing a binary search to find the maximum value of $\eta$. The algorithm for computing the robust bound $B$ for a specific paragraph is detailed in Algorithm~\ref{alg:boundcompute} in Appendix \ref{sec:app:pseudo_code}. The complexity of Algorithm~\ref{alg:boundcompute} is $O(n \cdot \eta_{max}^4 + t \cdot \log \eta_{max})$, dominated by the computation of the $\mathsf{PI}$ array. In practice, we typically use $n \leq 10$ and $\eta_{max} = 40$, allowing $B$ to be efficiently computed within a few seconds on a modern CPU.

%\vspace{-2em}

\revise{
\subsection{Security analysis}

In cryptography, hash functions are often modeled under the random oracle model (ROM)~\cite{ROM,ROM2}. This model is widely deployed for analysis in cryptography. The random oracle $H:\{0,1\}^* \rightarrow\{0,1\}^n$ is defined as follows:
\begin{itemize}
    \item For every input $x, H(x)$ produces a uniformly random $n$-bit output.
    \item The response is the same for identical queries, but is independent with other inputs.
\end{itemize}
We first analyze the theoretical security of our watermark's secret key under the random oracle model (ROM), which means that $\mathsf{hash}(sk,\cdot)$ is viewed as a random oracle. The output of the hash function cannot be directly observed by the watermark's secret key extraction adversary.  To simplify analysis, we consider a stronger adversary $\mathcal{A^*}$, formalized as:\\
\begin{itemize}
    \item $\mathcal{A^*}$ observes a set of pairs $\left\{\left(X_j, \mathsf{hash}\left(\mathrm{sk}, X_j\right)\right)\right\}_{j=1}^q$, where $X_j$ are adaptively chosen bit-strings, and $q$ is the number of queries.
    \item After $q$ queries, $\mathcal{A^*}$ outputs a candidate for $sk$.
\end{itemize}

Then, the problem becomes analyzing whether it is possible to extract the secret key by observing hash outputs.
Due to computation power limitation, assume $\mathcal{A^*}$ tests at most $Q$ key guesses, $\{sk'_1,\ldots, sk'_Q\}$. The test is done by checking $\mathsf{hash}(sk'_i,X_j)=\mathsf{hash}(sk,X_j), j=1,\ldots,q$. Due to the uniform randomness of ROM, the advantage of $\mathcal{A^*}$ of finding the exact $sk$ in $Q$ queries compared with completely random guessing is bounded by:
\begin{small}
    $$
\operatorname{Pr}[\mathcal{A^*} \text { succeeds }] - \operatorname{Pr}[\mathcal{A^*} \text { succeeds by random guessing}]\leq \frac{Q}{2^n}
$$
\end{small}

In contrast, if $\exists$ adversary $\mathcal{A^*}$  recover $sk$ with probability advantage significantly higher than $\frac{Q}{2^n}$, it indicates that $\mathcal{A^*}$ already leverages the structure of the output of $\mathsf{hash}$. This breaks the assumption that $\mathsf{hash}$ (viewed as a random oracle) provides independent random outputs. 

In practice, the output bit length of the hash function is usually $n\ge 128$. The computation power of $\mathcal{A^*}$ is assumed as $Q\le 2^{64}$. Thus the probability of $\mathcal{A^*}$ to succeed in extracting the secret key is no bigger than $\frac{Q+1}{2^n}\approx 2^{-64}$, which is negligible.

From the above analysis, we can see that it is very difficult to extract the secret key even by directly observing the hash outputs. Thus we prove that it is very difficult for an adversary to recover the secret key by observing watermarked text.

Next, we consider the security of our watermark against spoofing attacks. In the case of spoofing attack, the adversary has access to $k$ ($k \geq 1$) distinct accounts, each associated with a unique user ID $m_1, m_2, ..., m_k$. For simplicity of analysis, we assume only the userID is embedded. The adversary's objective is to forge a piece of text that will be considered as originating from the target user ID $m_t$ ($m_t \notin \{m_1,...,m_k\}$). 

Similarly, we also consider a stronger adversary $\mathcal{A'}$, who can adaptively query the hash function with user IDs among $m_1,...,m_k$ and observe the output. The target of the adversary is to guess the hash output of $\mathsf{hash}(sk,m_t)$ based on observations.

Based on the random oracle assumption, the hash results of different inputs are uniformly random. Therefore, $\forall m_t \neq m_i (i \in \{1, 2,...,k\})$, $\mathsf{hash}(sk,m_t)$ and $\mathsf{hash}(sk,m_i)$ are independent random variables. In other words, the adversary $\mathsf{A'}$ gain no advantage in predicting $\mathsf{hash}(sk,m_t)$ by querying $\mathsf{hash}(sk,m_i)$. Consequently, the hash value $\mathsf{hash}(sk,m_t)$ remains uniformly random from the adversary's perspective. As the actual adversary in spoofing attack is strictly weaker than the adversary $\mathcal{A'}$ (he cannot observe any actual hash values), our watermark is theoretically resistant to spoofing attack under ROM.
}

\section{Experiments}

In this section, we present a comprehensive set of experiments designed to rigorously evaluate the correctness, robustness and efficiency of our watermarking method for LLMs.

\subsection{Experiment setup}
\noindent
\textbf{Datasets.} Following~\cite{pmlr-v202-kirchenbauer23a,yoo2023advancing} on watermarking texts generated by LLMs, we utilize the following datasets for our evaluation: OpenGen~\cite{krishna2023paraphrasing}, C4 news dataset~\cite{raffel2020exploring}, and Essays dataset~\cite{essaysdataset}. In particular, OpenGen is composed of 3,000 two-sentence chunks randomly sampled from the WikiText-103~\cite{merity2016pointer} validation set. The C4 news dataset consists of 15GB news crawled from the internet. The essays dataset consists of homework essays written by students from the IvyPanda essay database~\cite{ivy2024panda}. Unless otherwise mentioned, we use the OpenGen dataset as it is the most popular one. 
%We utilize OpenGen~\cite{krishna2023paraphrasing} for our main experiments, following the previous work~\cite{zhao2023provable}. OpenGen is composed of 3K two-sentence chunks randomly sampled from the WikiText-103~\cite{merity2016pointer} validation set. For dataset ablation, we also adopt C4 news dataset ~\cite{raffel2020exploring} and Essays dataset ~\cite{essaysdataset}, following previous works ~\cite{pmlr-v202-kirchenbauer23a,yoo2023advancing}.

\noindent
\textbf{Large language models.}
We conduct our major experiments with state-of-the-art public LLMs: LLaMA-2-7B~\cite{touvron2023llama}, Guanaco-7B~\cite{dettmers2024qlora}, and Falcon-7B~\cite{almazrouei2023falcon}. By default, we use LLaMA-2-7B~\cite{touvron2023llama} as it is more popular than the other two.

\noindent
\textbf{LLM decoding algorithms.}
For the LLM decoding algorithm, we use greedy sampling by default. We consider Multinomial sampling and Nucleus sampling~\cite{holtzman2019curious} when studying the impact of decoding algorithms on our method.

\begin{figure*}[!t]
\centering
\subfloat[OpenGen]{\includegraphics[width=0.3\textwidth]{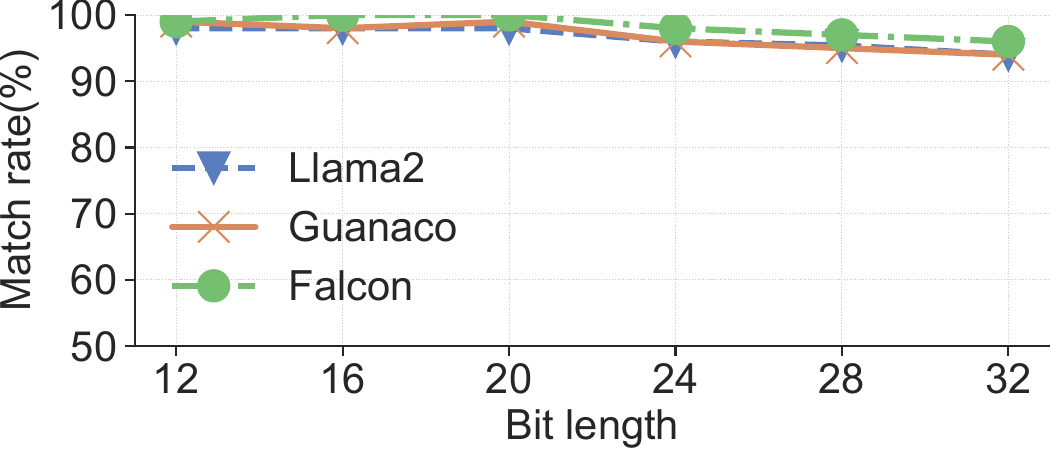}}
\subfloat[C4]{\includegraphics[width=0.3\textwidth]{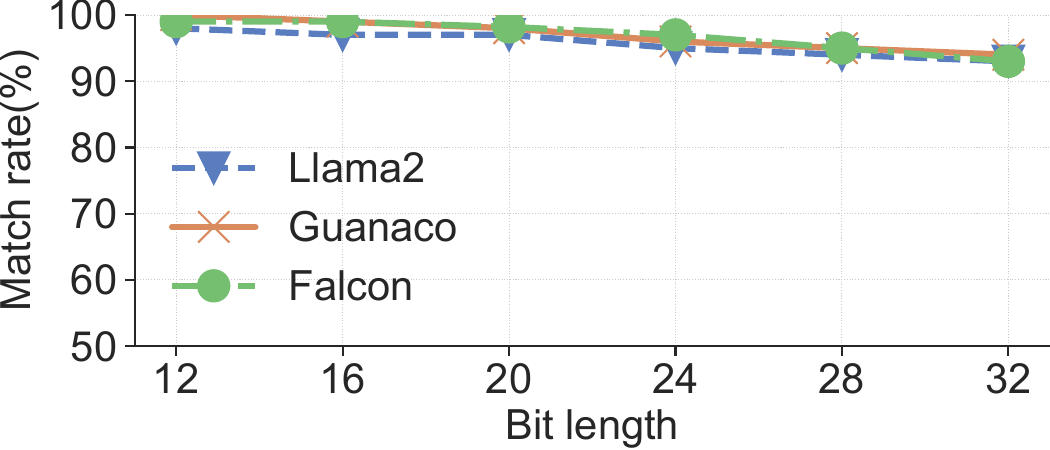}}
\subfloat[Essays]{\includegraphics[width=0.3\textwidth]{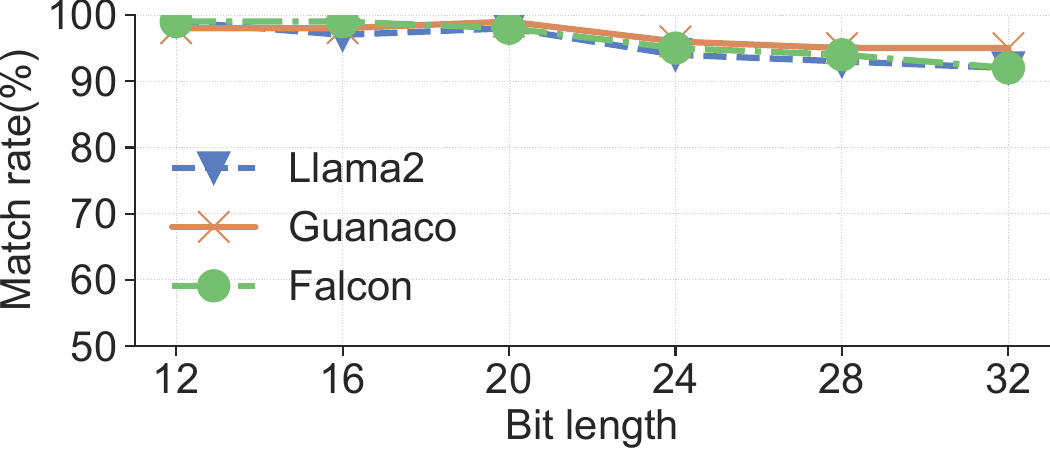}}

\caption{Match rate of our method on different datasets, LLMs, and bit lengths.}
%\vspace{-2em}
\label{fig:differentmatchrate}
\end{figure*}

\begin{figure*}[h]
\centering
\subfloat[Opengen]{\includegraphics[width=0.3\textwidth]{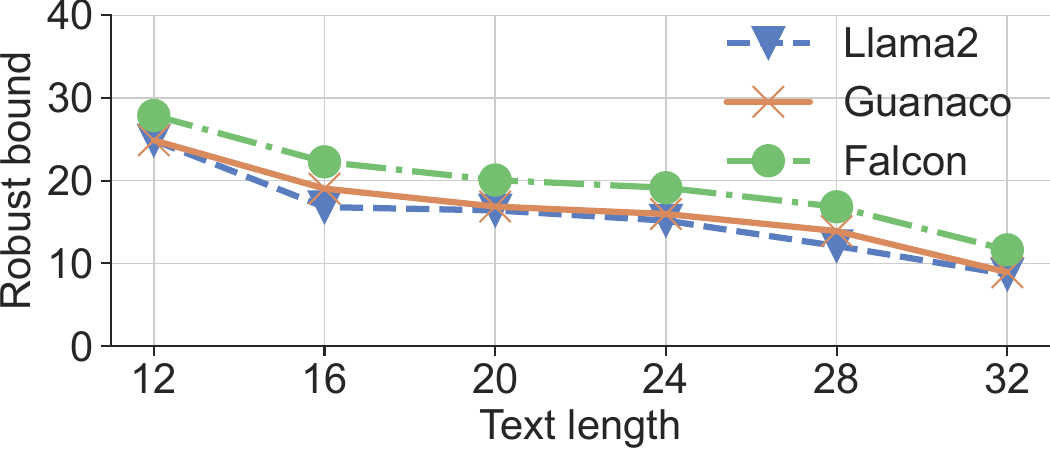}}
\subfloat[C4]{\includegraphics[width=0.3\textwidth]{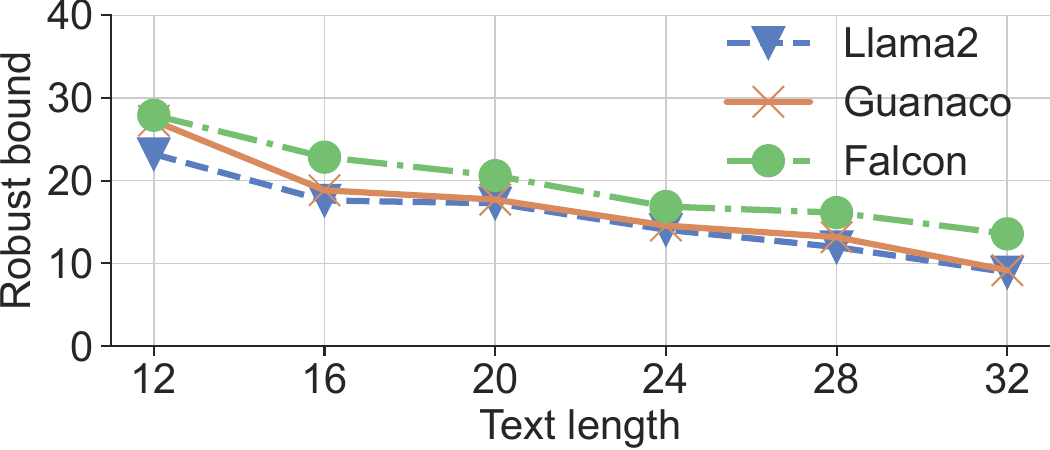}}
\subfloat[Essays]{\includegraphics[width=0.3\textwidth]{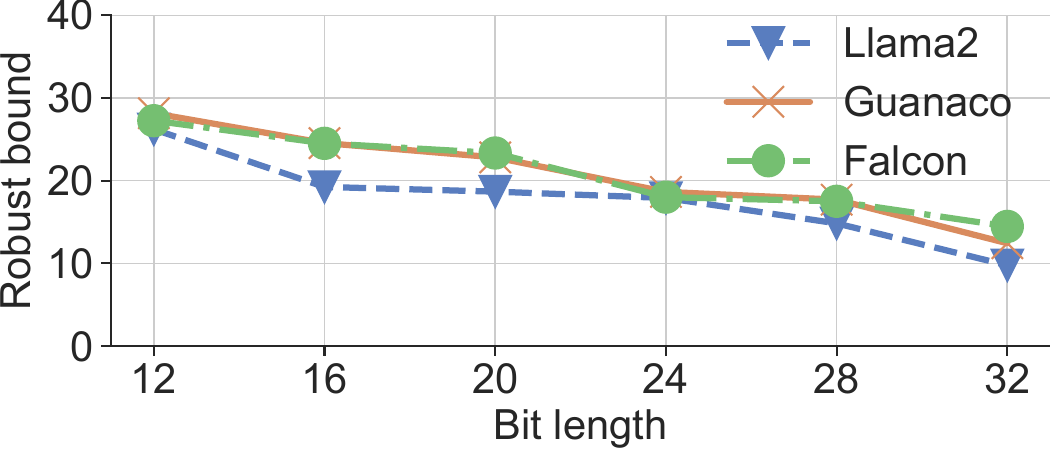}}

\caption{ Robust bound of our method on different datasets, LLMs, and bit lengths. }
\label{fig:differentrobust}
\end{figure*}
\noindent
\textbf{Metrics and evaluation methods. }
We mainly use \textbf{match rate} to measure the proportion of generated text that can exactly extract embedded watermark message without error. We also test bit accuracy (the ratio of correctly extracted message bits) which is adopted in several
previous work \cite{luo2020distortion, yang2022tracing, yoo2023robust, yoo2023advancing}. 
We use \textbf{edit distance} to quantify the attacker's change to the watermarked text. Edit distance (Levenshtein distance)~\cite{navarro2001guided} measures the difference between two texts defined as the minimum number of single-token edits (insertions, deletions, or substitutions) required to change one text into the other. Additionally, to measure theoretical robustness, we define the metric \textbf{robust bound} for a dataset as the average robust bound for each generated text. The robust bound for each paragraph is calculated using Algorithm~\ref{alg:boundcompute} in Appendix \ref{sec:app:pseudo_code}. We set the error probability threshold in Definition~\ref{define:robust} as $\alpha=10^{-3}$. For text quality, following the work in~\cite{pmlr-v202-kirchenbauer23a}, we use perplexity (PPL) computed by a large oracle model GPT-3.5 (text-davinci-003)~\cite{ouyang2022training}.  

\noindent
\textbf{Parameter settings. }
In our experiments, we adopt a default setting with bias $\delta = 6$. We investigate token generation for a max generation length of $T = 200$ tokens, embedding a 20-bit message , i.e., $b=20$, which we randomly sample for each paragraph. For selecting the Reed-Solomon code in Algorithm~\ref{alg:encode}, we determine a minimum code rate $R_c = 0.6$, minimum recover rate $R_r=0.15$. For the default setting of embedding $20$-bit message, we opt for the Reed-Solomon code scheme with  $(n, k, t)_{2^m} = (6,4,1)_{2^5}$.  Throughout this section, the term ``bit length'' refers to the length of the message that we intend to embed into a text using Algorithm~\ref{alg:encode}, i.e. $b$. We conduct extensive ablation studies to evaluate the impact of each hyperparameter on our method.

\noindent
\textbf{Hardware.}
All of our experiments are conducted on a server with 80 Intel Xeon @2.1GHz CPUs, and 4 V100-32GB GPUs.

\subsection{Main results}

\begin{table*}[h]

\caption{Compare our method with existing baselines, on match rate and extraction time (per text paragraph). \revise{Match rate and bit accuracy are shown in percentages. "NA" means running the corresponding experiment on the dataset would take a few weeks or months, so we decided not to proceed with it.}} 

\centering
\resizebox{\textwidth}{!}{%
\revise{
\begin{tabular}{|c|ccc|ccc|ccc|ccc|ccc|}
\hline
\multirow{2}{*}{Bit length $b$}   & \multicolumn{3}{c|}{12}                                                                                                                                                                                  & \multicolumn{3}{c|}{16}                                                                                                                                                                                  & \multicolumn{3}{c|}{20}                                                                                                                                                                                  & \multicolumn{3}{c|}{24}                                                                                                                                                                                  & \multicolumn{3}{c|}{32}                                                                                                                                                                                            \\ \cline{2-16} 
                 & \multicolumn{1}{c|}{\begin{tabular}[c]{@{}c@{}}Match \\ rate\end{tabular}} & \multicolumn{1}{c|}{\begin{tabular}[c]{@{}c@{}}Bit\\ Acc\end{tabular}} & \begin{tabular}[c]{@{}c@{}}Time\\ (s)\end{tabular} & \multicolumn{1}{c|}{\begin{tabular}[c]{@{}c@{}}Match \\ rate\end{tabular}} & \multicolumn{1}{c|}{\begin{tabular}[c]{@{}c@{}}Bit\\ Acc\end{tabular}} & \begin{tabular}[c]{@{}c@{}}Time\\ (s)\end{tabular} & \multicolumn{1}{c|}{\begin{tabular}[c]{@{}c@{}}Match \\ rate\end{tabular}} & \multicolumn{1}{c|}{\begin{tabular}[c]{@{}c@{}}Bit\\ Acc\end{tabular}} & \begin{tabular}[c]{@{}c@{}}Time\\ (s)\end{tabular} & \multicolumn{1}{c|}{\begin{tabular}[c]{@{}c@{}}Match \\ rate\end{tabular}} & \multicolumn{1}{c|}{\begin{tabular}[c]{@{}c@{}}Bit\\ Acc\end{tabular}} & \begin{tabular}[c]{@{}c@{}}Time\\ (s)\end{tabular} & \multicolumn{1}{c|}{\begin{tabular}[c]{@{}c@{}}Match \\ rate\end{tabular}} & \multicolumn{1}{c|}{\begin{tabular}[c]{@{}c@{}}Bit\\ Acc\end{tabular}} & \begin{tabular}[c]{@{}c@{}}Time\\ (s)\end{tabular}           \\ \hline
Fernandez et al.~\cite{fernandez2023three} & \multicolumn{1}{c|}{99.6}                                                  & \multicolumn{1}{c|}{100.0}                                                  & 0.12                                               & \multicolumn{1}{c|}{99.6}                                                  & \multicolumn{1}{c|}{100.0}                                                  & 0.34                                               & \multicolumn{1}{c|}{99.2}                                                  & \multicolumn{1}{c|}{99.9}                                                  & 5.04                                               & \multicolumn{1}{c|}{98.0}                                                  & \multicolumn{1}{c|}{99.8}                                                  & 110                                                & \multicolumn{1}{c|}{NA}                                                    & \multicolumn{1}{c|}{NA}                                                  & \begin{tabular}[c]{@{}c@{}}29000 \\ (Estimated)\end{tabular} \\ \hline
Wang et al.~\cite{wang2023towards} & \multicolumn{1}{c|}{99.6}                                                  & \multicolumn{1}{c|}{100.0}                                                  & 0.16                                               & \multicolumn{1}{c|}{98.8}                                                  & \multicolumn{1}{c|}{99.9}                                                  & 0.58                                               & \multicolumn{1}{c|}{98.4}                                                  & \multicolumn{1}{c|}{99.8}                                                  & 3.14                                               & \multicolumn{1}{c|}{97.2}                                                  & \multicolumn{1}{c|}{99.6}                                                  & 35.5                                               & \multicolumn{1}{c|}{NA}                                                    & \multicolumn{1}{c|}{NA}                                                  & \begin{tabular}[c]{@{}c@{}}8300 \\ (Estimated)\end{tabular}  \\ \hline
Yoo et al.~\cite{yoo2023advancing}       & \multicolumn{1}{c|}{86.4}                                                  & \multicolumn{1}{c|}{96.5}                                                  & 0.01                                               & \multicolumn{1}{c|}{73.6}                                                  & \multicolumn{1}{c|}{94.6}                                                  & 0.01                                               & \multicolumn{1}{c|}{49.2}                                                  & \multicolumn{1}{c|}{90.8}                                                  & 0.01                                               & \multicolumn{1}{c|}{30.4}                                                  & \multicolumn{1}{c|}{89.4}                                                  & 0.01                                               & \multicolumn{1}{c|}{8.4}                                                   & \multicolumn{1}{c|}{81.2}                                                  & 0.01                                                         \\ \hline
Cohen et al.~\cite{cohenwatermarking}      & \multicolumn{1}{c|}{93.2}                                                  & \multicolumn{1}{c|}{98.7}                                                  & 0.01                                               & \multicolumn{1}{c|}{88.8}                                                  & \multicolumn{1}{c|}{97.0}                                                  & 0.02                                               & \multicolumn{1}{c|}{78.4}                                                  & \multicolumn{1}{c|}{95.3}                                                  & 0.02                                               & \multicolumn{1}{c|}{65.6}                                                  & \multicolumn{1}{c|}{94.7}                                                  & 0.03                                               & \multicolumn{1}{c|}{27.2}                                                  & \multicolumn{1}{c|}{89.7}                                                  & 0.04                                                         \\ \hline
Ours             & \multicolumn{1}{c|}{98.8}                                                  & \multicolumn{1}{c|}{99.9}                                                  & 0.04                                               & \multicolumn{1}{c|}{98.0}                                                  & \multicolumn{1}{c|}{99.7}                                                  & 0.06                                               & \multicolumn{1}{c|}{97.6}                                                  & \multicolumn{1}{c|}{99.6}                                                  & 0.10                                               & \multicolumn{1}{c|}{96.0}                                                  & \multicolumn{1}{c|}{99.5}                                                  & 0.18                                               & \multicolumn{1}{c|}{94.0}                                                  & \multicolumn{1}{c|}{99.1}                                                  & 0.6                                                          \\ \hline
\end{tabular}
}

}
%\vspace{-3mm}
\label{tab:baseline}
\end{table*}

\begin{figure}[h]

\centering
\subfloat{\includegraphics[width=0.3\textwidth]{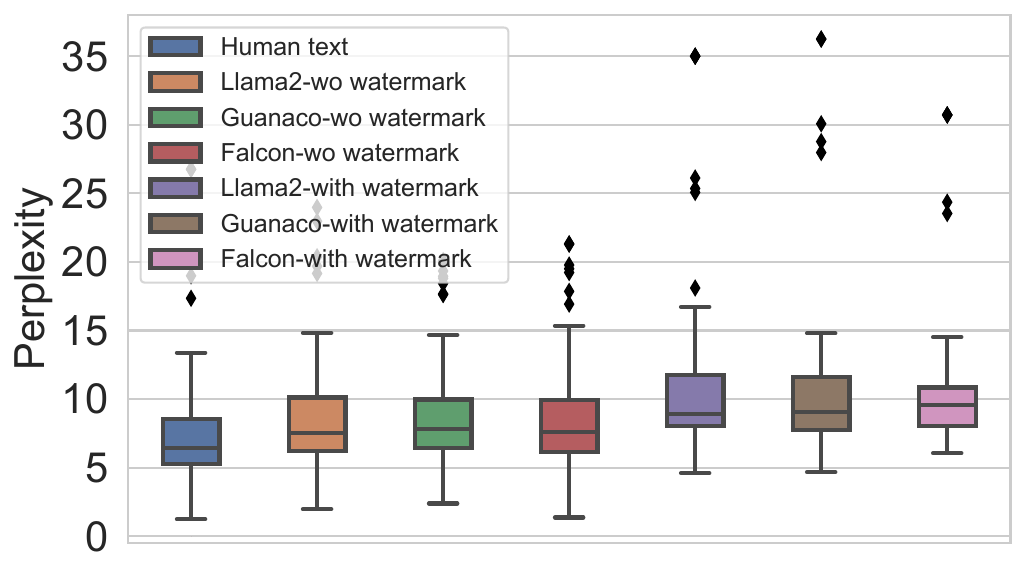}}
\caption{Our method maintains the quality of texts generated by LLMs. The perplexity of texts generated by LLMs is similar with and without our watermark. }
\label{fig:ppl_model}
\end{figure}

\begin{table*}[h]
\centering
\caption{The impact of different Reed-Solomon code parameters. The scheme adopted by our watermark is in red.}
\scalebox{0.77}{
\revise{
\begin{tabular}{|l|ccc|ccc|ccc|ccc|}
\hline
Bit length & \multicolumn{3}{c|}{16}                          & \multicolumn{3}{c|}{20}                          & \multicolumn{3}{c|}{24}                          & \multicolumn{3}{c|}{32}                          \\ \hline
RS Code    & \multicolumn{1}{c|}{ \textcolor{red}{$(6,4,1)_{2^4}$}  } & \multicolumn{1}{c|}{$(8,4,2)_{2^4}$ } & $(4,2,1)_{2^8}$ & \multicolumn{1}{c|}{ \textcolor{red}{$(6,4,1)_{2^5}$}   } &  \multicolumn{1}{c|}{$(8,4,2)_{2^5}$ } & $(7,5,1)_{2^4}$ & \multicolumn{1}{c|}{\textcolor{red}{$(5,3,1)_{2^8}$}} & \multicolumn{1}{c|}{$(7,3,2)_{2^8}$} & $(6,4,1)_{2^6}$  & \multicolumn{1}{c|}{\textcolor{red}{$(6,4,1)_{2^8}$}} & \multicolumn{1}{c|}{$(8,4,2)_{2^8}$} & $(10,8,1)_{2^4}$ \\ \hline
Match rate & \multicolumn{1}{c|}{\textcolor{red}{98.0}} & \multicolumn{1}{c|}{96.4} & 97.2 & \multicolumn{1}{c|}{\textcolor{red}{97.6}} & \multicolumn{1}{c|}{95.6} & 97.6 & \multicolumn{1}{c|}{\textcolor{red}{96.0}} & \multicolumn{1}{c|}{94.4} & 95.6 & \multicolumn{1}{c|}{\textcolor{red}{94.0}} & \multicolumn{1}{c|}{91.2} & 90.4 \\ \hline
\end{tabular}
}
}
\label{tab:rs_code}
\end{table*}

\begin{table}[h]
\centering
\revise{
\caption{\revise{User study on evaluating watermarked text quality.}}
\begin{tabular}{|l|c|c|}
\hline
              & Avg Score & STD   \\ \hline
Unwatermarked & 3.816     & 0.543 \\ \hline
Watermarked   & 3.844     & 0.562 \\ \hline
\end{tabular}
\label{tab:user_study}
}
\end{table}

\noindent
\textbf{Our method achieves the correctness goal. } 
To assess the adaptability of our method under diverse settings, we carry out experiments using three models—Llama2, Guanaco, and Falcon—and three datasets: OpenGen, C4, and Essays. We evaluate bit lengths of 12, 16, 20, 24, 28, and 32 while keeping other hyperparameters at their default settings. The results are depicted in Figure \ref{fig:differentmatchrate}. It is observed that the match rates for a given bit length are relatively consistent across different models and datasets, demonstrating the versatility of our watermarking method. Notably, there is a discernible trend where the match rate diminishes as the bit length increases. This trend is natural as when the bit length grows, the information redundancy available for embedding each bit becomes less. 
%With fewer assigned tokens for one segment, the likelihood of its count of green tokens for the correct value being less than other candidates increases. 
In summary, the proposed solution has yielded remarkable results. In all cases, we have attained a match rate exceeding 90\% for all three datasets.

\noindent
\textbf{Our method achieves the robustness goal. } 
We evaluate the theoretical robustness of our method across different settings. We conduct the experiment under the same setting as the previous correctness experiment. The results are illustrated in Figure \ref{fig:differentrobust}. We find that the theoretical robustness bound for different models and different datasets are also quite similar under the same bit length.
The results also show that, as the bit length increases, the robust bound decreases. This result is not surprising, as it is rooted in theoretical analysis. When the bit length increases, the number of ECC-coded segments $n$ or the segment size $m$ also increases. Consequently, the RHS of Inequality~\ref{formula:robust_prob_RHS} escalates with the increase of $n$ or $m$. When $n$ or $m$ increases, under the same $\eta$, $p_{0}^{(n)}$ defined in Section~\ref{sec:theory_robust} decreases. According to Theorem \ref{thm:robust_theorem}, the RHS of Inequality~\ref{formula:robust_prob_RHS} rises with the increase of $n$ or $m$. Therefore, the robust bound
%, defined as the maximum $\eta$ for which RHS $\le \alpha$, 
will diminish with the increase of bit length.

%This presents a challenge in developing a watermark that maintains strong provable robustness while embedding a longer bit string in a text comprising 200 tokens. 

\noindent
\textbf{Our method maintains the quality of texts generated by LLMs. }
We also explore the influence of our watermark on the quality of text generated by an LLM. Experiments are conducted on OpenGen dataset using 3 different models—Llama2, Guanaco, Falcon, and different bit lengths of 12, 16, 20, 24, 28, and 32. Other hyperparameters are under the default settings. Figure~\ref{fig:ppl_model} uses box plots to compare the perplexity distribution of human-written text with that of texts generated by three different models (Llama2, Guanaco, and Falcon), both with and without watermarking. Note that a lower perplexity indicates higher text quality.  The results indicate that watermarking only slightly harms the quality of AI-generated text, but the effect is minor. Prior to the application of our watermark, the perplexity values associated with text generated by Falcon predominantly fall within the interval of $[6,10]$. Following the integration of our watermark, a very slight elevation in perplexity occurs, with the generated text now primarily falling within the range of $[6.5,10.5]$. The results suggest that our watermarking method subtly affects the quality of the generated text, ensuring the utility of watermarked AI-generated content in practical applications.

\revise{
We also conduct a user study to assess the text quality following~\cite{zhao2023provable}.  We generate 100 watermarked text and 100 unwatermarked text using LLaMA2-7B model on the OpenGen dataset, and ask 25 participants to rate the text quality on
a 5-points Likert Scale from 1 (poor) to 5 (excellent). Each generated text receives 2 rated scores.  The average score and standard
deviation are presented in Table~\ref{tab:user_study}. The results indicate that, the text quality degradation caused by the watermark is also subtle according to human evaluation.
}

\noindent
\textbf{Our method outperforms baselines \cite{yoo2023advancing, fernandez2023three,wang2023towards}.}
 To compare our approach with baselines \cite{yoo2023advancing, fernandez2023three,wang2023towards,cohenwatermarking}, we measure match rate and extraction time with bit lengths set to 12, 16, 20, 24, 32 separately, while other parameters are kept at their default values. We do not compare with ~\cite{boroujeny2024multi} since they are not open-sourced. Additionally, their method is not practical when embedding 20-bit messages, see our analysis in Appendix~\ref{app:analyze_distortion_free}.  
 
 %We do not compare with ~\cite{cohenwatermarking} because it is a fully theoretical scheme and does not conduct any experiments or release code \wenjie{need discuss}. 
 
 The experimental results for baseline comparisons are shown in Table \ref{tab:baseline}. For match rate, our method achieves results similar to those reported in Fernandez et al.~\cite{fernandez2023three}, especially at larger bit lengths, where they were significantly better than the findings of Yoo et al.~\cite{yoo2023advancing}. In terms of extraction time, our results are very close to those of \cite{yoo2023advancing}. 
 For~\cite{fernandez2023three,wang2023towards}, the extraction time increases exponentially with the increase of bit length, because their methods require to enumerate over  $2^{b}$ possible messages. For a bit length of $b = 24$, our method achieves a match rate of 96.0\%, markedly surpassing the 30.4\% achieved by \cite{yoo2023advancing}. Additionally, our extraction process consumes less than 1 second, in contrast to the extraction time exceeding 100 seconds as reported by~\cite{fernandez2023three}. For bit length $b=32$, we do not run the entire experiment for~\cite{fernandez2023three,wang2023towards} due to their infeasible time overhead.

\begin{figure*}[!t]
\centering
\includegraphics[width=0.85\textwidth]{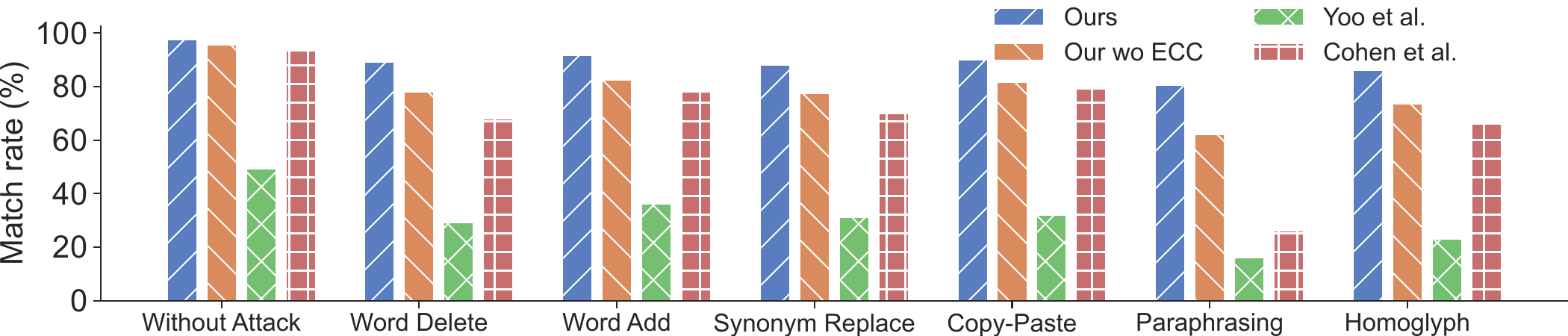}

\caption{A comparison of our watermark algorithm with Yoo et al.\cite{yoo2023advancing}, Cohen et al.~\cite{cohenwatermarking}, and our algorithm without error-correction code under different attacks.}
\label{fig:empirical_robust}

\end{figure*}

\noindent
\revise{
    \textbf{Impact of Reed-Solomon code parameters.} To demonstrate the effectiveness of our ECC scheme selection algorithm, we evaluate different RS code parameters under different bit lengths. The results are presented in Table~\ref{tab:rs_code}. Across the bit lengths of 16, 20, 24, 32,  our selected RS code consistently achieves the highest match rate for our watermark. It can also be observed that increasing the number of recovery bits often fails to improve the match rate. This is because although more recovery bits increase the recover ability of ECC, it also results in larger $n$ which exacerbates the imbalance in token assignment. Although our algorithm generally selects the ECC with the smallest $n$, it does not consistently do so. For example, for bit length $16$, RS code $(6,4,1)_{2^4}$ chosen by our algorithm  is better than $(4,2,1)_{2^8}$. This is because compared to embedding $6\times4=24$ bits with $(6,4,1)_{2^4}$, $(4,2,1)_{2^8}$ requires embedding $4\times8=32$ bits into generated text, resulting in insufficient text entropy to encode all 32 bits, thereby increasing the likelihood of extraction errors.  In conclusion, our ECC code selection algorithm simultaneously considers the number of segments $n$, the code rate, and recovery rate to achieve the best performance.
}

%\noindent
%\textbf{Impact of ECC scheme selection strategy.}
%Table \ref{tab:eccselection}  \ref{appendix: experiment result} effectively showcases the results of our ECC scheme selection process, which we've designed to ensure that each watermark carries enough information while maintaining a balance between redundancy and error recovery. Through our methodology, we have identified ECC configurations that achieve an optimal balance for various bit lengths. For a bit length of 12, we observe that the (23,12,3) scheme achieves an impressive match rate of 98.4\%. It is higher than that achieved by the (31,16,3) scheme, which has a higher redundancy. This outcome validates our approach of imposing a minimum code rate $R$. When the bit length is increased to 16, the (31,16,3) scheme provides the best trade-off, since it offers the highest recovery rate among the candidates, resulting in the highest match rate. Similarly, for bit length 20, the (41,21,4) scheme stands out with a high match rate of 93.2\%, showcasing our method's scalability with increased bit lengths. The configurations highlighted in red underline the efficiency of our ECC selection process, ensuring robust watermarking across a range of text lengths.

\subsection{Robustness results}

In the real world, a user may attempt to edit the generated text to evade the watermark or improve the utility of the text. We investigate six types of attacks previously studied in the literature~\cite{pmlr-v202-kirchenbauer23a, yoo2023advancing, zhang2023remark}: (1) Word Deletion Attack~\cite{zhang2023remark} (2) Word Addition Attack (3) Synonym Replacement Attack~\cite{zhang2023remark}  (4) Paraphrasing Attack~\cite{zhang2023remark}
(5) Copy-paste Attack~\cite{yoo2023advancing}
(6) Homoglyph Attack~\cite{pmlr-v202-kirchenbauer23a}. Following ~\cite{yoo2023advancing,zhang2023remark}, we set attack types (1) (2) (3) (5) to change 10\% of the tokens.
%: modify text by replacing characters with the same or a very similar-looking letter. For example, for the word "hello" (one token), the attacker replaces the character "l" with its similar-shaped Unicode characters, such as "1". 
We compare our algorithm with Yoo et al.~\cite{yoo2023advancing}, Cohen et al.~\cite{cohenwatermarking} and our algorithm without error-correction code under these attacks. We do not compare with Fernandez et al.~\cite{fernandez2023three} and Wang et al.~\cite{wang2023towards} here due to their inefficiency as shown in Table~\ref{tab:baseline}.

In our paper, we consider an attacker who aims to distort the watermark while preserving the semantics and readability of the resulting texts. Therefore, we do not consider the emoji attack proposed in ~\cite{pmlr-v202-kirchenbauer23a}, as it compromises readability. Furthermore, the attacker employs LLMs to expedite text generation. Consequently, we do not consider scenarios involving the manual rewriting of texts for watermark removal, as the additional human effort deviates from the attacker's objectives of automating text generation.

Figure~\ref{fig:empirical_robust} shows the results of the attack experiments. We observe that our algorithm still persists high accuracy compared with other baselines under different attacks. For example, under the copy-paste attack, our method achieves a match rate of $90.0\%$, while Yoo et al.~\cite{yoo2023advancing} only achieves $32.4\%$. And under the word deletion attack, our method significantly outperforms Yoo et al.~\cite{yoo2023advancing} by $60\%$. 
These results clearly demonstrate that our method achieves strong robustness under different attacks. The major reason for our method achieving much better robustness compared to Yoo et al.~\cite{yoo2023advancing} is our design of segment assignment. Segment assignment reduces the probability of extracting incorrect messages by mitigating imbalance. 

%Reed-Solomon code offers our scheme the ability to tolerate error segments and substantially improves robustness.

The experimental results also demonstrate the importance of our design of adopting error-correction code. For example, under homoglyph attack, our method achieves an accuracy of $86.4\%$, while not adopting ECC can only achieve $73.6\%$. Adopting ECC also contributes to about 10\% match rate gain under the synonym replacement attack and word deletion attack.  This robustness advantage is even more significant under the most challenging paraphrasing attack. Adopting 
 ECC improves match rate by 18\%, and the match rate of our method still achieves 80\% under paraphrasing attack. 
The rationale behind ECC improving robustness is as follows: These attacks reduce the number of green tokens in each group, leading to incorrect values being extracted from some segments. By adopting ECC, we can tolerate errors in some segments during extraction, which significantly increases the difficulty of distorting our embedded message.

\section{Conclusion}
In this work, we proposed a watermarking method tailored for AI-generated texts. Our method 
 achieves four goals simultaneously: correctness, robustness, multi-bit capacity, and efficiency. The effectiveness of our method lies in the design of pseudo-random segment assignment and two improvement techniques.  Furthermore, we establish the provable robustness of our watermarking method against adversarial edits. Empirical results highlight our watermarking method's superior performance in both correctness and efficiency compared to existing methods.
 
\section{Acknowledgment} We would like to extend our gratitude to the reviewers and shepherd for their constructive feedback on our work. We are also grateful to Polyhedra for providing the computing resources necessary for our experiments. Additionally, we thank Jian Liu, Zixin He for their valuable feedback.
 
 %Future work could be conducted in designing multi-bit watermarks which work for  text that is very short.
 
 %Addressing the unsolved challenge of embedding long bit strings in short text could advance the use of multi-bit watermarks in practical applications. 
 %While acknowledging a limitation in embedding long bit strings into relatively short text pieces, overcoming this constraint could signify a pivotal stride towards deploying multi-bit watermarks in practical applications.

\bibliographystyle{plainurl}
\bibliography{reference}

\appendix

\section{Proof of Robustness Theorem~\ref{thm:robust_theorem}}
\label{appendix:proof}

We provide the high level ideas of our proof and the formal proof of our robustness theorem in this section. The proof relies on the following assumption.   
\begin{assumption}
Assume $hash(\cdot)$ is a pseudo-random function, and the attacker cannot predict $hash(x)$ given an arbitrary input $x$.
\label{ass:random}
\end{assumption}

\paragraph{Proof Sketch.} 
Consider a scenario where the attacker's modifications to the text \( S \) are confined within a predetermined edit distance. Such alterations result in bounded changes to the \(\mathsf{COUNT}\) matrix. We define a `wrong segment' as a segment that decodes incorrectly. By leveraging the properties of the ECC scheme characterized by parameters \((n, k, t)_{2^m}\), our objective is to demonstrate that the probability of the \(\mathsf{COUNT}\) matrix having more than \(t\) wrong segments is smaller than a limited probability \(\alpha\). This probability is effectively bounded as per the right-hand side (RHS) of Inequality~\ref{formula:robust_prob_RHS}. To compute the RHS, we first need to compute \( p_i^{(n)} \). We derive \( p_i^{(n)} \) by considering the cases where the \(n\)-th segment is extracted correctly and incorrectly. Subsequently, \( p_i^{(n)} \) can be computed based on \( p_{i-1}^{(n-1)} \) and \( p_{i-1}^{(n)} \), forming a recurrence relation for calculating \( p_i^{(n)} \). The proof details are as follows.

\begin{proof} 

The following notations are crucial for our proof.
\begin{definition}
Denote $\mathsf{B}(n,p)$ as the binomial distribution, where $\Pr(X=k|X\sim\mathsf{B}(n,p))=\binom{n}{k}p^k(1-p)^{n-k}$.
\end{definition}

\begin{definition}
Denote $\mathsf{H}(N,K,n)$ as the hypergeometric distribution, where $\Pr(X=k|X\sim\mathsf{H}(N,K,n))=\frac{\binom{K}{k}\binom{N-K}{n-k}}{\binom{N}{n}}$.
\end{definition}

%We first consider the simple case where $g=T-1$.\\
\begin{definition}
    Define $\Bigram(S)=\{(S_0,S_1),(S_1,S_2),\cdots,(S_{T-2},S_{T-1})\}$.
\end{definition}

We claim that each operation of insertion, deletion, and edition  can remove at most two existing elements and add at most two elements in $\Bigram(S)$. We consider insertion, deletion, and edition separately. \\

For insertion at position $i$, the token $S*$ is inserted. $(S_{i-1},S_i)$ is removed while $(S_{i-1}, S*),(S*, S_{i})$ are added.
For deletion at position $i$, the token $S_i$ is deleted. $(S_{i-1},S_i)$, $(S_i,S_{i+1})$ are removed while $(S_{i-1}$, $S_{i+1})$ are added.
For edition at position $i$,  the token $S_i$ is changed to $S*$. $(S_{i-1},S_i)$, $(S_i,S_{i+1})$ are removed while $(S_{i-1}, S*)$,$(S*, S_{i+1})$ are added.\\

Therefore, after at most $\eta$ such operations, the attacker can remove at most $2\eta$ existing elements and add at most $2\eta$ existing elements in $\Bigram(S)$. This indicates that for $D(S, S') \le \eta$, the number of tuples added $x$ and the numbers of tuples deleted $y$ are both bounded by $2\eta$.
%We assume the attacker removes $2\eta$ existing elements, and adds $2\eta$ new elements.\\

Using error-correction code $(n, k, t)_{2^m}$, we can correct at most $t$ incorrect segments. Based on the definition of $p_i^{(k)}$, to compute the probability of correct extraction, we can sum the corresponding probabilities $p_i^{(n)}(x,y,\textbf{c},\textbf{d})$ for $i\in\{0,1,\cdots,t\}$. Based on the analysis of $\Bigram(S)$, we have:
\begin{align}
    \begin{aligned}
       \Pr(\mathsf{Extraction}(S')\neq K) \le 1-\sum_{i=0}^t p_i^{(n)}(2\eta,2\eta,\textbf{c},\textbf{d})
   % p_i^{(n)}(x,y)
    \end{aligned}
    \label{eq:core}
\end{align}

Since we leveraged $p_i^{(k)}$ to derive an upper bound for the probability of incorrect extraction, now the challenge is how to derive and compute $p_i^{(k)}$.  We start from defining and analyzing the case of a single segment.

\begin{definition}
Define $f(x,y,c,d)$ as the probability that a segment is decoded correctly, given its allocated token number $c$, green token number $d$, number of tokens  $x$ added to the corresponding group  and number of tokens $y$ deleted from the group. 
\end{definition}

Now we derive the probability expression of $f(x,y,c,d)$. This function illustrates the probability of a single segment decoding correctly after a certain number of token changes. Editing causes some allocated tokens to be removed and some tokens to be added.

Let $X$ be the added green token number of the correct value. For each added token, the probability that it is a green token of the correct value is $\frac{1}{2}$. For $x$ added tokens, we have $X\sim\mathsf{B}(x,\frac{1}{2})$.

Let $Y$ be the deleted green token number of the correct value. For deleting $y$ tokens from a set of $c$ tokens, among which $d$ is the green token number of the correct value, we have $Y\sim\mathsf{H}(c,d,y)$.

Let $d'$ be the green token number of the correct value after adding x tokens and deleting y tokens. We have $d'=d+X-Y$.

%Let $Z$ be the green token number of a wrong value after adding $x$ tokens and deleting $y$ tokens. For each wrong value, we have $Z\sim\mathsf{B}(a',\frac{1}{2})$.

$f(x,y,c,d)$ can be derived as follows:
\begin{align}
    \begin{aligned}
       & f(x,y,c,d)=\\
&\sum_{d'=0}^{d+x}\Pr(d+X-Y=d'|X\sim\mathsf{B}(x,\frac{1}{2}),Y\sim\mathsf{H}(c,d,y))\\
& \cdot\Pr(d' >  2^m-1 ~\text{value candidates} | \text{initially~} c ~\text{tokens},\\
&   ~\text{added~} x ~\text{tokens~and~deleted~}y~\text{tokens} )
    \end{aligned}
    \label{define_f}
\end{align}

Under Assumption~\ref{ass:random}, the green token number of each incorrect value candidate can be viewed as sampled from $\mathsf{B}(c+x-y,\frac{1}{2})$. %The green token number of each pair's second value is $c+x-y$ minius the first value's green token number. Therefore, 
%\begin{small}
    \begin{align}
    \begin{aligned}
   & \Pr(d' > ~\text{one incorrect value's green token number} |\\
&  \text{initially~} c ~\text{tokens,~added~} x ~\text{tokens~and~deleted~}y~\text{tokens} )\\
    &    =1-\Pr(Z\ge d'|Z\sim\mathsf{B}(c+x-y,\frac{1}{2}))
    \end{aligned}
\end{align}
%\end{small}

There are in total $2^{m}-1$ incorrect values. The green list of these values are independent with each other. Thus, we have:\\
\begin{align}
    \begin{aligned}
&\Pr(d' >  2^m-1 ~\text{value candidates} | \text{initially~} c ~\text{tokens},\\
&   ~\text{added~} x ~\text{tokens~and~deleted~}y~\text{tokens} )\\
&    =(1-\Pr(Z\ge d'|Z\sim\mathsf{B}(c+x-y,\frac{1}{2})))^{2^{m}-1}
    \end{aligned}
    \label{eq:bprime}
\end{align}

Substituting Equation~\ref{eq:bprime} into Equation~\ref{define_f}, we have:\\
\begin{align}
    \begin{aligned}
       & f(x,y,c,d)=\\
&\sum_{d'=0}^{d+x}\Pr(d+X-Y=d'|X\sim\mathsf{B}(x,\frac{1}{2}),Y\sim\mathsf{H}(c,d,y))\\
&(1-\Pr(Z\ge d'|Z\sim\mathsf{B}(c+x-y,\frac{1}{2})))^{2^{m}-1}
    \label{compute_f}
    \end{aligned}
\end{align}

After deriving function $f$ illustrating the decoding probability distribution of a single segment after editing, we consider the distribution of the decoding correctness of multiple segments illustrated by $p_i^{(k)}(x,y,\textbf{c},\textbf{d})$.
We derive the recursive formula to compute $p_i^{(k)}(x,y,\textbf{c},\textbf{d})$. The probability density function (PDF) of $p_i^{(k)}$ is crucial for computing our final robust bound.

We start from the special case of  $k=1,i=0$.
It is obvious that $p_0^{(1)}(x,y,\textbf{c},\textbf{d})$ is the probability of the first segment decoding to the correct value. We have: 
\begin{align}
    \begin{aligned}
        p_0^{(1)}(x,y,\textbf{c},\textbf{d})=f(x,y,c_1,d_1)
    \end{aligned}
\end{align}

Additionally, $p_1^{(1)}(x,y,\textbf{c},\textbf{d})$ represents the probability of the complementary event of the previous event:

\begin{align}
    \begin{aligned}
        p_1^{(1)}(x,y,\textbf{c},\textbf{d})=1-p_0^{(1)}(x,y,\textbf{c},\textbf{d})
    \end{aligned}
\end{align}

Denote the number of tokens added to the $k$-th segment's group as $X'$. For adding $x$ tokens to the first $k$ segments, we have $X'\sim\mathsf{B}(x,\frac{1}{k})$.

Denote the number of tokens deleted from the $k$-th segment's group as $Y'$. For deleting $y$ tokens from $(c_1,\cdots,c_n)$ tokens, we have $Y\sim\mathsf{H}(\sum_{j=1}^kc_j,c_k,y)$.

For the probability $p_i^{(k)}(x,y,\textbf{c},\textbf{d})$ that exactly $i$ segments in the first $k$ segments decode incorrectly, if $i\ge 1$, there are two cases: the first case is that the $k$-th segment decodes incorrectly and exactly $i-1$ segments in the first $k-1$ segments decode incorrectly; the second case is that the $k$-th segment decodes correctly and exactly $i$ segments in the first $k-1$ segments decode incorrectly.

The probability of the first case is $p_{i-1}^{(k-1)}(x-x_k,y-y_k,\textbf{c},\textbf{d})
(1-f(x_k,y_k,c_k,d_k))$.

The probability of the second case is $p_i^{(k-1)}(x-x_k,y-y_k,\textbf{c},\textbf{d}) f(x_k,y_k,c_k,d_k)$.

Therefore, for $i\ge 1$, we have the recursive formula for $p_i^{(k)}(x,y,\textbf{c},\textbf{d})$:\\
%\begin{tiny}
%\begin{footnotesize}

\begin{align}
    \begin{aligned}
    &p_i^{(k)}(x,y,\textbf{c},\textbf{d})\\
    &=\sum_{x_k=0}^x\Pr(X=x_k|X\sim\mathsf{B}(x,\frac{1}{k}))\sum_{y_k=0}^y\Pr(Y=y_k|Y\sim\mathsf{H}(\sum_{j=1}^k c_j,c_k,y))\\
&( p_{i-1}^{(k-1)}(x-x_k,y-y_k,\textbf{c},\textbf{d})
(1-f(x_k,y_k,c_k,d_k))+\\
&p_i^{(k-1)}(x-x_k,y-y_k,\textbf{c},\textbf{d})f(x_k,y_k,c_k,d_k))        
    \end{aligned}
    \label{eq:recurs1}
\end{align}
%\end{tiny}
%\end{footnotesize}

If $i=0$, the only possibility is all the first $k$ segments decode correctly. Similarly, we have:
%\begin{footnotesize}
\begin{align}
    \begin{aligned}
    &p_0^{(k)}(x,y,\textbf{c},\textbf{d})\\
    &=\sum_{x_k=0}^x\Pr(X=x_k|X\sim\mathsf{B}(x,\frac{1}{k}))\sum_{y_k=0}^y\Pr(Y=y_k|Y\sim\mathsf{H}(\sum_{j=1}^k c_j,c_k,y))\\
&p_0^{(k-1)}(x-x_k,y-y_k,\textbf{c},\textbf{d})f(x_k,y_k,c_k,d_k)        
    \end{aligned}
    \label{eq:recurs2}
\end{align}
%\end{footnotesize}

In conclusion, we can leverage Equation~\ref{eq:recurs1} and Equation~\ref{eq:recurs2} to recursively compute every $p_i^{(k)}$ for $i\le t, k\le n$. By substituting each $p_i^{(n)}$ in Equation~\ref{eq:core}, we obtain an upper bound of the probability of error extraction after editing the text within an edit distance of $\eta$.

% For text paragraph $S$ generated by Algorithm \ref{alg:encode}. Denote embedding information $K\in \{0,1\}^{km}$ with token number $T$, error-correction code $(n, k, t)_{2^m}$, initial token number for each symbol $(a_1,\cdots,a_n)$ and initial green token number for each symbol $(b_1,\cdots,b_n)$. Attacker modifies $S$ within edit distance budget $\eta$ to obtain $S'$ such that $D(S, S') \le \eta$. 

\end{proof}

\begin{figure}[h]
%\vspace{-5mm}
\centering
\subfloat{\includegraphics[width=0.4\textwidth]{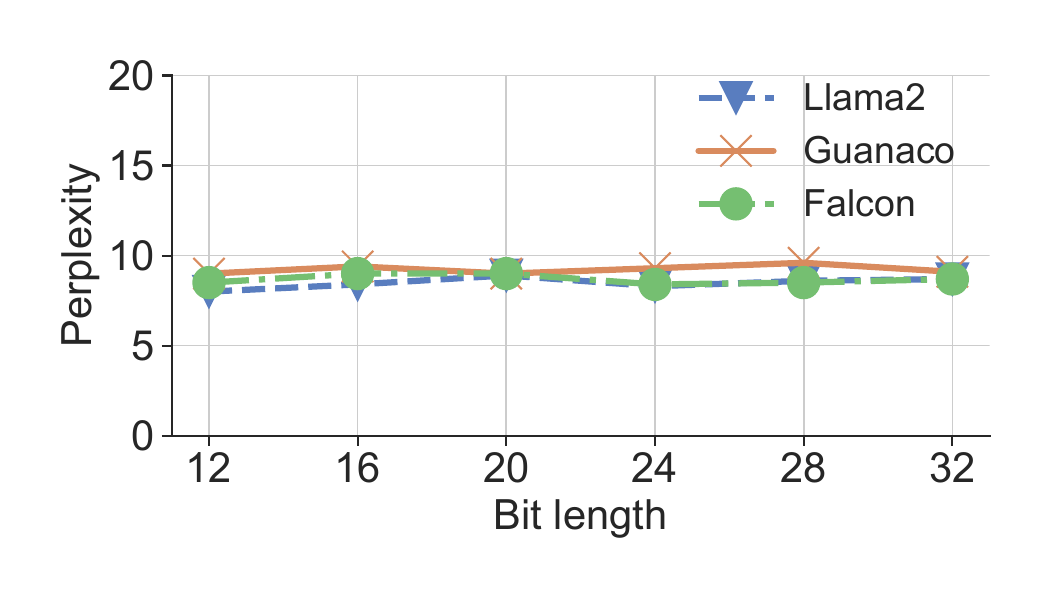}}
\caption{The perplexity of texts generated by LLMs with our watermark is similar across different bit lengths and models.}
\label{fig:ppl_bit_length}

\end{figure}

%\begin{table}[h]
%\centering
%\caption{Results of embedding bit string with large length. Bit length $b=64$. }
%\begin{tabular}{|c|c|c|c|c|}
%\hline
%Token number T &200 & 300  & 400  & 500  \\ \hline
%Match rate~(\%)    &26.4 & 67.2 & 84.0 & 92.0 \\ %\hline
%\end{tabular}
%\label{tab:bitlength32}
%\end{table}

%\begin{table}[h]
%\centering
%\caption{Match rates under different Reed-Solomon code schemes.\wenjie{need data update.}}
%\vspace{-3mm}
%\begin{tabular}{|c|c|c|}
%\hline
%Bit length $b$          & ECC scheme            & Match rate            \\ \hline
%\multirow{3}{*}{12} & (21,12,2)             & 97.2                  \\ \cline{2-3} 
%                   & 
%\textcolor{red}{(23,12,3)}  &      98.4                 \\ \cline{2-3} 
%                    & (31,16,3)             & 96.8                  \\ \hline
%\multirow{3}{*}{16}          & (31,21,2)  & 82.4                      \\ \cline{2-3} 
%                & 
%\textcolor{red}{(31,16,3)}  &  96.7                   \\ \cline{2-3} 
%                & (33,23,2)  & 78.0                       \\         
 %              \hline
%\multirow{3}{*}{20}          & (31,21,2)  &     82.4                  \\ \cline{2-3} 
 %               & 
%\textcolor{red}{(41,21,4)}  &    93.2                  \\ \cline{2-3} 
 %               & (47,24,5)  &    86.8                    \\         
  %             \hline
%\end{tabular}
%\label{tab:eccselection}
%\end{table}

\section{Pseudo-code of our procedures}
\label{sec:app:pseudo_code}
Algorithm~\ref{alg:choose_rs} illustrates our procedure of selecting the optimal Reed-Solomon code scheme.
%Algorithm~\ref{alg:encode_pair} illustrates our procedure for embedding watermark with pairing.
%Algorithm~\ref{alg:decode_pair} illustrates our procedure for extracting watermark with the message embedded with segment value pairing.
Algorithm~\ref{alg:boundcompute} illustrates our algorithm for computing the provable bound for each watermarked paragraph.
Algorithm~\ref{alg:pik_compute} illustrates our algorithm for computing the $\mathsf{PI}$ array used in Algorithm~\ref{alg:boundcompute}.

\begin{algorithm}[tb]
   \caption{\emph{Select Reed-Solomon scheme}}
   \label{alg:choose_rs}
\begin{algorithmic}
    \STATE {\bfseries Input:} set of Reed-Solomon scheme $\mathcal{F}$; embed bit number $b$; minimal code rate $R_c$; minimal recovery rate $R_r$
    \STATE {\bfseries Output:} Optimal Reed-Solomon scheme $(\hat{n}, \hat{k}, \hat{t})_{2^{\hat{m}}}$
    \STATE Initialize candidate schemes set $\mathcal{C}$ as empty set 
    \FOR{$(n,k,t)_{2^m} \in \mathcal{F}$}
    \IF{$k\cdot m = b$, code rate $\frac{k}{n} \geq R_c$, recover rate $\frac{t}{n} \geq R_r$}
    \STATE $\mathcal{C}$ = $\{(n,k,t)_{2^m}\} \bigcup \mathcal{C}$
    \ENDIF
    \ENDFOR

    \STATE $(\hat{n}, \hat{k}, \hat{t})_{2^{\hat{m}}} \leftarrow (n,k,t)_{2^m} \in \mathcal{C}$ with minimal $n$
    \STATE \textbf{return}  $(\hat{n}, \hat{k}, \hat{t})_{2^{\hat{m}}}$
\end{algorithmic}
\end{algorithm}

\begin{algorithm}[tb]
   \caption{\emph{Compute robust bound for watermarked text}}  
   \label{alg:boundcompute}
\begin{algorithmic}
   \STATE {\bfseries Input:} Text token number $T$; Reed-Solomon scheme $(n,k,t)$; initial
allocated token numbers for each segment $\mathbf{c}$; initial green token numbers for each segment $\mathbf{d}$; robust edit distance upper bound $\eta_{max}$; maximum error rate $\alpha$\\
   \STATE {\bfseries Output:} Robust bound for text: $ \eta$\\
    \STATE $\eta_{1}=0$\\
    \STATE $\eta_{2}=\eta_{max}$\\
    \STATE Initialize array $\mathsf{PI}[t+1][2\eta_{max}+1][2\eta_{max}+1]$\\
    \STATE Compute $\mathsf{PI}[i][x][y]=p_i^{(n)}(x,y,\textbf{c},\textbf{d})$ for $i\in [0,t], x,y\in [0,2\eta_{max}]$ using Algorithm~\ref{alg:pik_compute}\\
    \WHILE{$\eta_2-\eta_1>1$}
    \STATE $\eta_{mid}=\lfloor \frac{\eta_1+\eta_2}{2} \rfloor$\\
    \STATE Initialize $P_E$ as 1\\
    \FOR{$i = 0, 1,\cdots, t$}
        %\STATE Compute $p_i^{(n)}(2\eta_{mid},2\eta_{mid},\textbf{c},\textbf{d})$ leveraging Algorithm~\ref{alg:pik_compute}. 
        \STATE $P_E \mathrel{-}= \mathsf{PI}[i][2\eta_{mid}][2\eta_{mid}]$
    \ENDFOR
    \IF{$P_E\le \alpha$}
    \STATE $\eta_1=\eta_{mid}$\\
    \ELSE
    \STATE $\eta_2=\eta_{mid}$\\
    \ENDIF
    \ENDWHILE
    \STATE \textbf{return} $\eta_1$\\

\end{algorithmic}
\end{algorithm}

\begin{algorithm}[tb]
\caption{\emph{Compute $p_i^{(n)}(x,y,\mathbf{c},\mathbf{d})$ table}}
\label{alg:pik_compute}
\begin{algorithmic}
\STATE {\bfseries Input:} Total segments number $n$; robust edit distance upper bound $\eta_{max}$; initial allocated token numbers for each segment $\mathbf{c}$; initial green token numbers for each segment $\mathbf{d}$ \\
\STATE {\bfseries Output:} $p_i^{(n)}(x,y,\mathbf{c},\mathbf{d})$ for $i\in[0,t]\  x,y\in [0,2\eta_{max}]$
\STATE Initialize all \( p_i^{(k)} \) to zero for \( k = 1, 2, \ldots, n \) and \( i = 0, 1, \ldots, t \)
\STATE $f \leftarrow $ Precomputed result table using Equation~\ref{compute_f}
\FOR{$x = 0, 1, \ldots, 2\eta_{max}$}
    \FOR{$y = 0, 1, \ldots, 2\eta_{max}$}
\STATE $p_0^{(1)}(x,y,\mathbf{c},\mathbf{d}) = f(x,y,c_1,d_1)$
\STATE $p_1^{(1)}(x,y,\mathbf{c},\mathbf{d}) = 1 - p_0^{(1)}(x,y,\mathbf{c},\mathbf{d})$
    \ENDFOR
\ENDFOR
\FOR{$k = 2, \ldots, n$}
    \FOR{$x = 0, 1, \ldots, 2\eta_{max}$}
    % \STATE \textbf{parallel for} $x = 0, 1, \ldots, 2\eta_{max}$ \textbf{do} \label{parfor:x_k}
        \FOR{$y = 0, 1, \ldots, 2\eta_{max}$}
            \FOR{$x_k = 0, 1, \ldots, x$}
                \FOR{$y_k = 0, 1, \ldots, y$}
                    \STATE $p_{x_k} = \Pr(X = x_k | X \sim \mathsf{B}(x, \frac{1}{k}))$
                    \STATE $p_{y_k} = \Pr(Y = y_k | Y \sim \mathsf{H}(\sum_{j=1}^k c_j, c_k, y))$

                    \STATE $p_0^{(k)}(x,y,\mathbf{c},\mathbf{d}) \mathrel{+}= p_{x_k} \cdot p_{y_k} \cdot (p_0^{(k-1)}(x-x_k, y-y_k, \mathbf{c}, \mathbf{d}) \cdot f(x_k, y_k, c_k, d_k))$
                    \FOR{$i = 1, \ldots, t$}
                        \STATE $p_i^{(k)}(x,y,\mathbf{c},\mathbf{d}) \mathrel{+}= p_{x_k} \cdot p_{y_k} \cdot (p_{i-1}^{(k-1)}(x-x_k, y-y_k, \mathbf{c}, \mathbf{d}) \cdot (1 - f(x_k, y_k, c_k, d_k)) + p_i^{(k-1)}(x-x_k, y-y_k, \mathbf{c}, \mathbf{d}) \cdot f(x_k, y_k, c_k, d_k))$
                    \ENDFOR
                \ENDFOR
            \ENDFOR
        \ENDFOR
    \ENDFOR
    % \STATE \textbf{end parallel for}
\ENDFOR
\STATE \textbf{return} $p_i^{(n)}(x,y,\mathbf{c},\mathbf{d})$ table\\
\end{algorithmic}
\end{algorithm}

\section{Analysis of Boroujeny et al.~\cite{boroujeny2024multi}}

\label{app:analyze_distortion_free}

Through our examination,~\cite{boroujeny2024multi} only works well when the bit length of message is small. With the bit length growing, this method has increasing difficulty in distinguishing between neighboring messages. 

Suppose the bit length of the message to embed is $b$. For any neighboring message $M = m$ and $M' = m+1$, and any binarized token $i$ with the probability of being 1 as $p_i(1)$, we have $\delta_M = M\delta = m \cdot 2^{-b}$ and $\delta_{M'} = M'\delta = (m+1)\cdot 2^{-b}$. Without losing generalizability, we first assume $p_i(1) + 2^{-b}(m+1) \leq 1$. According to Equation (21) in~\cite{boroujeny2024multi}, we have:

\begin{footnotesize}
\begin{align}
\label{multi_distort_free_interval}
\begin{aligned}    
&A_{1,i}(M) = \{0 \leq y_i < m \cdot 2^{-b}\} \bigcup \{p_i(1) + m \cdot 2^{-b} \leq y_i \leq 1 \} \\
&A_{2,i}(M) = \{m \cdot 2^{-b} \leq y_i < p_i(1) + m \cdot 2^{-b}\} \\
&A_{1,i}(M') = \{0 \leq y_i < (m+1) \cdot 2^{-b}\} \bigcup \{p_i(1) + (m+1) \cdot 2^{-b} \leq y_i \leq 1 \} \\
&A_{2,i}(M') = \{(m+1) \cdot 2^{-b} \leq y_i < p_i(1) + (m+1) \cdot 2^{-b}\} \\
\end{aligned}
\end{align}
\end{footnotesize}

According to Algorithm 3 in~\cite{boroujeny2024multi}, the $i$-th binarized token generated with message $M = m $ (denoted as $w_i^M$) and with message $M' = m+1$ (denoted as $w_i^{M'}$) will be

\begin{align}
\begin{aligned}
w^M_i = \vmathbb 1 [y_i \in A_{2,i}(M)] \\
w^{M'}_i = \vmathbb 1 [y_i \in A_{2,i}(M')]
\end{aligned}
\end{align}

To make $w^M_i \neq w^{M'}_i$, we need $((y_i \in A_{2,i}(M)) \land (y_i \notin A_{2,i}(M'))) \lor ((y_i \notin A_{2,i}(M)) \land (y_i \in A_{2,i}(M')))$. According to \ref{multi_distort_free_interval}, only when $y_i \in [m \cdot 2^{-b}, (m+1) \cdot 2^{-b}) \bigcup [p_i(1) + m \cdot 2^{-b}, p_i(1) + (m+1) \cdot 2^{-b})$ will $w^M_i \neq w^{M'}_i$ hold. As $y_i$ is sampled uniformly from $[0,1]$, the probability of sampling different binarized tokens for neighboring messages at any given position $i$  is $((m+1) \cdot 2^{-b} - m \cdot 2^{-b}) + (p_i(1) + (m+1) \cdot 2^{-b} - p_i(1) - m \cdot 2^{-b}) = 2^{1-b}$.

For conditions where $p_i(1) + 2^{-b}(m+1) > 1$, it can be readily confirmed that the probability is also $2^{1-b}$.

Based on the description in~\cite{boroujeny2024multi}, one normal token will be divided into 17 binarized tokens. Therefore, the probability of generating exactly the same $k$-tokens text under neighboring message $M$ and $M'$ is $(1-2^{1-b})^{17 \times k}$.

When the bit length $b$ increases to over 20 bits for practical usage, the probability of 2 neighboring messages generating exactly the same long text will become extremely high. Take bit length $b = 20$ and token length $k = 200$ as an example. The probability of 2 neighboring messages generating exactly the same 200-tokens text will become $(1 - 2^{-19})^{17 \times 200} \approx 0.9935$. Even if we already know the correct message is either $M$ or $M'$, as these 2 different messages will generate exactly the same 200 tokens text with extremely high probability, we cannot effectively discern whether the message is $M$ or $M'$ given the text no matter which method we use.

%%%%%%%%%%%%%%%%%%%%%%%%%%%%%%%%%%%%%%%%%%%%%%%%%%%%%%%%%%%%%%%%%%%%%%%%%%%%%%%%
\end{document}